\documentclass[referee]{aa} % for a referee version
\usepackage{graphicx}
\begin{document}
\def\lsim{\,\lower2truept\hbox{${< \atop\hbox{\raise4truept\hbox{$\sim$}}}$}\,}
\def\gsim{\,\lower2truept\hbox{${> \atop\hbox{\raise4truept\hbox{$\sim$}}}$}\,}

   \title{Trade-off between angular resolution 
and straylight contamination in 
%cosmological frequency channels of 
CMB anisotropy experiments.}

   \subtitle{II. Straylight evaluation.}

   \author{C.~Burigana\inst{1}
          \and
          M.~Sandri\inst{1}
          \and
          F.~Villa\inst{1}
          \and 
          D.~Maino\inst{2}
          \and 
          R.~Paladini\inst{3}
          \and 
          C.~Baccigalupi\inst{3}
          \and 
          M.~Bersanelli\inst{2}
          \and
          N.~Mandolesi\inst{1}
          }

   \offprints{C. Burigana}

   \institute{IASF/CNR, Sezione di Bologna, 
              via P.~Gobetti, 101, I-40129 Bologna, Italy\\
              \email{~burigana,~sandri,~villa,~mandolesi~@bo.iasf.cnr.it}
             \and 
             Dipartimento di Fisica,
             Universit\'a degli Studi di Milano, 
             via Celoria, 16, I-20133 Milano, Italy \\
             \email{~bersanelli,~maino~@uni.mi.astro.it}
             \and 
             SISSA, International School for Advanced Studies,
             via Beirut, 2-4, I-34014 Trieste, Italy \\
             \email{~paladini,~bacci~@sissa.it} \\
             \\
             On behalf of the LFI Consortium
             }

   \date{Sent March 27, 2003}
%\date{Received September 15, 1996; accepted March 16, 1997}

   \abstract{Satellite CMB anisotropy missions, such as WMAP
and {\sc Planck}, and also the new generation of balloon-borne and 
ground experiments, make use of complex multi-frequency instruments
at the focus of a meter class telescope to allow the 
joint study of CMB and foreground anisotropies, necessary for a
high quality component separation. 
%In the so-called ``cosmological window'', 
Between $\sim 70$~GHz and $\sim 300$~GHz, where 
foreground contamination is minimum, it is extremely 
important to reach the best trade-off between 
the improvement of the angular resolution, necessary to 
measure the high order acoustic peaks of CMB anisotropy, and the
minimization of the straylight contamination mainly due to the 
Galactic emission.
% (GSC, Galaxy Straylight Contamination).
This is one of the most critical systematic effects at large and 
intermediate angular scales 
(i.e. at multipoles $\ell$ less than $\approx 100$)
and constists in unwanted radiation entering the beam at large
angles from the direction of the antenna boresight direction.
We focus here, as a working case, on the 30 and 100~GHz channels
of the {\sc Planck} Low Frequency Instrument (LFI).
By assuming the nominal {\sc Planck} scanning strategy, 
we evaluate the GSC introduced by the most relevant
Galactic foreground components 
for a reference set of optical configurations, accurately 
simulated as described in Paper~I. We show that it is possible 
to improve the angular resolution of $5-7$~\% by keeping 
the overall peak-to-peak GSC below the level of few $\mu$K 
(and about 10 times smaller in terms of {\sc rms}).
%as necessary to avoid systematic errors comparable with
%the {\sc Planck} sensitivity. 
A comparison between 
the level of straylight introduced by the different Galactic components
for different beam regions (intermediate and far sidelobes)
is presented. Simple approximate relations giving the {\sc rms} and 
peak-to-peak levels of the GSC for the intermediate pattern 
and far sidelobes as functions of the corresponding contributions 
to the integrated antenna pattern are provided.
We compare the results obtained at 100~GHz
with those at 30~GHz, where GSC is more critical. Finally, 
for some reference cases we compare 
the results based on Galactic foreground templates derived from 
radio and IR surveys with those based on WMAP maps including
CMB and extragalactic source fluctuations.
%This paper is based on {\sc Planck} LFI activities.

   \keywords{Cosmology: cosmic microwave background -- Galaxy: general -- Space vehicles --
Telescopes -- Methods: data analysis.}
   }

\authorrunning{C. Burigana et al.}
\titlerunning{Trade-off between angular resolution and straylight contamination. II}

   \maketitle
%
%________________________________________________________________

\section{Introduction}

After the detection of cosmic microwave background 
(CMB) anisotropies at few degree scales by COBE/DMR
(Smoot et al. 1992, Bennett et al. 1996, G\'orski et al. 1996) and
the recent balloon-borne and ground experiments
(see Bersanelli et al. 2002 and references therein 
for a review on the pre-WMAP observational status)
%(De~Bernardis et al. 2000, Hanany et al. 2000)
at high sensitivity and 
resolution on limited sky regions
probing a universe model with $\Omega_{tot} \sim 1$
(see e.g. Netterfield et~al. 2002, Stompor et~al. 2001,
Pryke et~al. 2002, and references therein)
%(Lange et al. 2000, Balbi et al. 2000, Jaffe et al. 2000),
the NASA space mission WMAP (Wilkinson Microwave Anisotropy Probe,
see Bennett et al. 2003a) 
derived the CMB anisotropy angular power spectrum 
with unprecedent sensitivity and reliability (Hinshaw et al. 2003b, Kogut et al. 2003)
and improved the accuracy in the determination of 
the most important cosmological parameters (Spergel et al. 2003). 

Future fundamental progresses in CMB anisotropy and polarization 
will be based on the {\sc Planck} mission 
by ESA~\footnote{http://astro.estec.esa.nl/Planck/} 
(Bersanelli et al. 1996, Tauber 2000, Villa et al. 2003), 
planned to be launched in the year 2007.

In particular, the Low Frequency Instrument (LFI, Mandolesi et al. 1998; see also 
Mandolesi et al. 2002) and
the High Frequency Instrument (HFI, Puget et al. 1998; see also
Lamarre et al. 2002) on-board {\sc Planck}
will cover together a wide frequency range (30--900 GHz) which
should significantly improve the accuracy of the subtraction
of foreground contamination from the primordial CMB anisotropy,
providing at the same time a gold mine
of cosmological as well as astrophysical information 
(see e.g. De~Zotti et al. 1999 and references therein).

To fully reach these scientific goals,
great attention has to be devoted
to properly reduce and/or subtract all the 
possible systematic effects.

The effect of optical distortions, both in the main beam and 
in the near and far sidelobes, has been widely recognized as 
one of the most critical systematics both in balloon experiments 
and in space missions (Page et al. 2003, Barnes et al. 2003). 

The systematic effect introduced by main beam distortions (Burigana et al. 1998)
can be in part reduced by adopting aplanatic configurations for the primary
mirror (Villa et al. 1998, Mandolesi et al. 2000a, Villa et al. 2002) 
and its effect on the CMB anisotropy power spectrum recovery, relevant 
at the multipoles of the acoustic peaks,
can be partially removed through dedicated deconvolution codes (Arnau et al. 2002)
provided that the main beam shape can be accurately reconstructed in flight
(Burigana et al. 2002).

The requirement on the rejection of unwanted radiation coming from directions far
from the optical axis (straylight) is stringent for {\sc Planck} and does not pertain only
the telescope itself, but the
entire optical system, including solar panels, shielding,
thermal stability and focal assembly components.
The variations of the spurious straylight signal 
introduce contaminations in the anisotropy measurements.
The removal of this effect in data analysis is in principle much more complicated 
than the subtraction of main beam distortion effect. This is due to 
the difficulty to accurately known the ``real'' antenna pattern 
(with both at ground and in flight reconstruction) 
at very low response levels.
% and also because the straylight signal, 
%when reduced through optical optimization, is expected 
%to be embedded in the true signal,
%noise and other systematical effects
%%, being at the same time not negligible
%(see e.g. Burigana et al. 2001).

The antenna response features far from the beam centre
(sidelobes) are determined largely by
diffraction and scattering from the edges of the mirrors and from nearby
supporting structures. 
Therefore, they can be reduced by decreasing the illumination at
the edge of the primary, i.e. increasing the edge taper 
(ET if expressed in dB; the linear edge taper, LET,
is $10^{-{\rm ET}/10}$),
defined as the ratio of the power per unit area
incident on the centre of the mirror to that incident on the edge.
Of course, the higher is edge taper, the lower is the sidelobe level
and the straylight contamination.
On the other hand, increasing the edge taper
has a negative impact on the angular resolution
for a fixed size of the primary mirror 
(see e.g. Mandolesi et al. 2000b).

In the ``cosmological window'' between $\sim 70$~GHz and $\sim 300$~GHz, where 
foreground contamination is minimum, it is extremely 
important to reach the best trade-off between 
the improvement of the angular resolution, necessary to 
measure the high order acoustic peaks of CMB anisotropy, and the
minimization of the straylight contamination due to the 
Galactic emission (GSC, Galaxy Straylight Contamination),
one of the most critical systematic effects, most relevant at large and 
intermediate angular scales 
(i.e. at multipoles $\ell$ less than $\approx 100$).

In this work we focus, as a working case, on the 100~GHz channels 
of {\sc Planck} Low Frequency Instrument, although the methods 
and the basic results described here can be applied also to other
{\sc Planck} frequency channels and to 
different CMB anisotropy experiments.
We shall also compare our results with simulated data at 30~GHz.

We will use here the detailed analysis on optical computations 
and the main optical results concerning the full antenna
pattern response given in Sandri et al. 2003 (hereafter Paper~I).
We present extensive simulations of the GSC due to 
Galactic foreground components relevant at 100~GHz and, at this purpose, 
we have made use of a wide set of simulated optical configurations in order to find
the best compromise between resolution and GSC therefore 
defining the maximum range of multipoles accessible to the considered
frequency channel and the level of straylight signal affecting
the data. 

In Sect.~2 we briefly describe the basic recipes to simulate
{\sc Planck} observations (see also Appendix~A), 
the adopted optical input, 
and maps of Galactic components. The results of the simulated
straylight contamination are described in Sect.~3, while the comparison 
between the results obtained for different antenna patterns and 
foreground components is presented in Sect.~4 (and in Appendix~B
for the comparison with a straylight simulation at 30~GHz).
Finally, we discuss the results and draw our main conclusions 
in Sect.~5.

\section{Simulations}

The selected orbit for {\sc Planck} is a Lissajous orbit
around the Lagrangian point L2 of the Sun-Earth system
(see e.g. Mandolesi et al. 1998).
The spacecraft spins at 1 r.p.m.
and the field of view of the two instruments (LFI/HFI) is 
about $10^\circ \times 10^\circ$ centered at the 
telescope optical axis (the so-called telescope line of sight, LOS) 
at a given angle $\alpha$ from the spin-axis direction,
given by a unit vector, $\vec s$, chosen to be pointed 
in the opposite direction with respect to the Sun.
In this work we consider values of  $\alpha \sim 85^{\circ}$,
as adopted for the baseline scanning strategy.
The spin axis will be kept parallel to the Sun--spacecraft direction
and repointed by $\simeq 2.5'$ every $\simeq 1$~hour (baseline
scanning strategy).
Hence {\sc Planck} will trace large circles in the sky.
A precession
of the spin-axis with a period, $P$, of $\simeq 6$~months at
a given angle $\beta \sim 10^{\circ}$
about an axis, $\vec f$, parallel to the Sun--spacecraft direction
(and outward the Sun) and shifted of $\simeq 2.5'$ every $\simeq 1$~hour,
may be included in the scanning strategy, possibly with 
a modulation of the speed of the precession in order to optimize
data transmission (Bernard et al. 2002). 
Although the scanning strategy could be changed, 
the GSC pattern, peak-to-peak, and angular power spectrum
are very weakly dependent on the details
of these proposed scanning strategies
(Burigana et al. 2000).

The code we have implemented for simulating
{\sc Planck} observations for a wide set of scanning
strategies is described in detail in Burigana et al. (1997, 1998)
and in Maino et al. (1999).
In this study we exploit the baseline scanning strategy 
and simply assume {\sc Planck} located in L2.
We do not consider the {\sc Planck} Lissajous orbit around L2
because its effects are negligible in this context.

We compute the convolutions between the antenna pattern response
and the sky signal as described in Burigana et al. (2001)
by working at $\sim 1^{\circ}$ or $\sim 7'$ resolution 
when considering the far or the intermediate sidelobes and by 
considering spin-axis shifts of $\sim 2^{\circ}$ every two days
and 180 samplings per scan circle. 
In fact, the effects of pattern features we want to study here occur
at $\sim$ degree or larger scales and, also, the wide set 
of optical simulations for the full antenna pattern is available 
at $\sim$ degree resolution
(see Paper~I for details on computation time of optical simulations).

With respect to the reference frames described in 
Burigana et al. (2001), following the recent developments 
in optimizing the polarization properties of LFI main beams
(see Paper~I), the conversion between the 
standard Cartesian {\it telescope frame} $x,y,z$ and the 
{\it beam frame} $x_{bf},y_{bf},z_{bf}$ requires 
a further angle $\psi_B$ other than
the standard polar coordinates $\theta_B$ and $\phi_B$
defining the colatitude and the longitude of the main beam 
centre direction in the {\it telescope frame}.
Appendix A provides the transformation rules 
between the {\it telescope frame} and the {\it beam frame}, as well
as the definition of the reference frames adopted in this work. 

The orientation of these frames as the satellite moves is implemented in
the code. For each integration time, we determine the orientations
in the sky of the {\it telescope frame} and of the {\it beam frame},
thus performing a direct convolution between the full pattern response 
and the sky signal for the desired number of maps, simultaneously.

\subsection{Optical inputs}

A detailed discussion of the optical simulation method and results
is presented in Paper~I (see Tables~2 and 3 of Paper~I for the main properties of 
the adopted antenna patterns).
We briefly summarize here the most relevant aspect in this context.

Several full beam patterns have been simulated for different
designs of two {\sc Planck} LFI feedhorns at 100~GHz differently located
on the focal plane unit: three models (9A, 9B, 9C) 
for LFI9 and four models (4A, 4B, 4C, 4D) for LFI4. 
Different values of ET will reflect in significant 
differences in the level of the GSC in the TOD,
while different feed designs with the same ET will produce
small, but not negligible, differences. 

The details of the antenna pattern response,
computed as described in Paper~I as functions of 
the two standard polar coordinates 
$\theta_{bf}$, $\phi_{bf}$ in the {\it beam frame},
depend also on the optical contributions considered 
in the analysis.
For some representative cases, we will compare the results of our 
simulations of GSC by adopting 
optical computations taking into account
the first and second or the first, second and third 
order optical interactions (see Paper~I). 
%(diffractions and/or reflections) 
%of rays with the two reflectors and the baffle of the 
%{\sc Planck} telescope in order to compare the impact of including 
%or neglecting the third interactions on the straylight analysis.

In the {\it beam frame} we identify two different angular regions 
relevant for the straylight analysis: the intermediate pattern 
and the far sidelobes, respectively
defined as the region between $1.2^\circ$ and $5^\circ$  from the
beam centre direction and at angles large than $5^\circ$ from the 
beam centre direction. The definition of these ``reference''
angles is, of course, somewhat arbitrary. They roughly separate angular
regions where significant pattern variations occur on sub-degree scales
from those where they occur on scales of few degrees or larger.
The values adopted here allow a direct comparison with a previous 
analysis (Burigana et al. 2001).

\subsection{Maps of Galactic components}

In the cosmological window (70--300~GHz) the CMB is clearly the dominant
component; on the other hand the Galactic emission is still relevant at 
low and middle Galactic latitudes. The templates adopted here are
similar to those described in Maino et al. (2002) and Paladini et al. (2003).

At 100~GHz the most relevant Galactic component is the thermal dust emission. 
We adopted here a template obtained by
extrapolating the maps by Schlegel et al. (1998) which combines
IRAS and DIRBE data, assuming a grey-body spectrum (expressed in antenna
temperature),
\begin{equation}
\label{dustscaling}
T_{A,dust}(\nu)\propto {\tilde{\nu}^{\beta+1}\over e^{\tilde{\nu}}-1}\ ,\
\tilde{\nu}={h\nu\over kT_{dust}}
\end{equation}
with uniform temperature $T_{dust}=18\,$K and emissivity $\beta=2$.

In order to simulate the free-free
contribution we assume, somewhat arbitrarily, that it is perfectly
correlated with the dust itself, i.e. that it has the same spatial
distribution. Its antenna temperature scales with frequency as
$T_{A,ff}\propto \nu^{-\beta_{ff}}$, with $\beta_{ff}=-2.1$. The
relative amplitude of dust and free-free emission is assumed to be
a factor of 3 at 100~GHz (De Zotti et al. 1999). Thus, 
we produce a single map of ``thermal'' emission from dust plus 
diffuse free-free emission (see Fig.~1)
with a spectrum described by:

\begin{equation}
\label{thermalscaling} T_{A,thermal}(\nu )=\left[{ {1\over 3}\left({\nu
\over 100{\rm GHz}}\right)^{\beta_{ff}}+ {T_{A,dust}(\nu )\over
T_{A,dust}(100 {\rm GHz})} }\right] \times T_{A,dust}(100 {\rm GHz}) \, .
\end{equation}

The synchrotron emission template is the 408 MHz map of Haslam et al.
(1982), available at a resolution of $0.85^{\circ}$, 
extrapolated to the considered frequencies (see Fig.~2) assuming a
uniform spectral index $ \beta_{\rm syn}=-2.9$ in antenna temperature.
No attempt is made here to add small scale fluctuations, since 
the effects of pattern features on which this work is focused
occur at $\sim$ degree or larger scales; for the same reason,
the fact the original template includes a convolution 
with a beam with the given resolution is not a concern.

   \begin{figure*}[t]
   \centering
   \includegraphics[width=10cm]{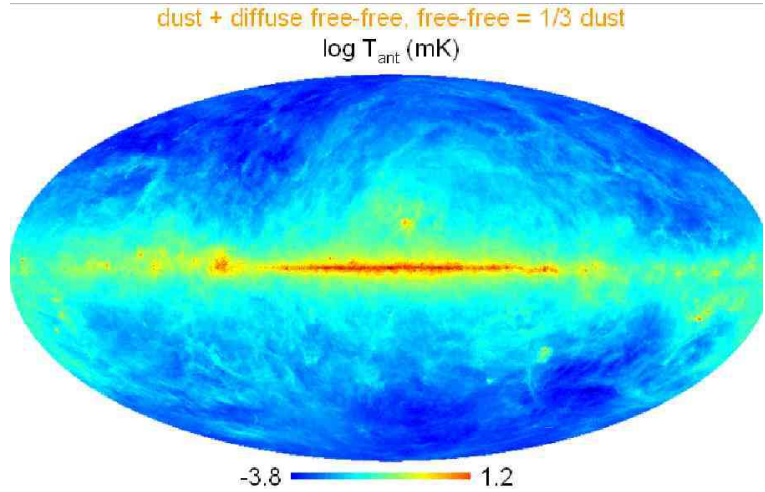}
   \caption{Map of the Galactic dust emission plus diffuse free-free emission
adopted in this study (see also the text).}
%              \label{FigGam}%
    \end{figure*}
   \begin{figure*}[!h]
   \centering
   \includegraphics[width=10cm]{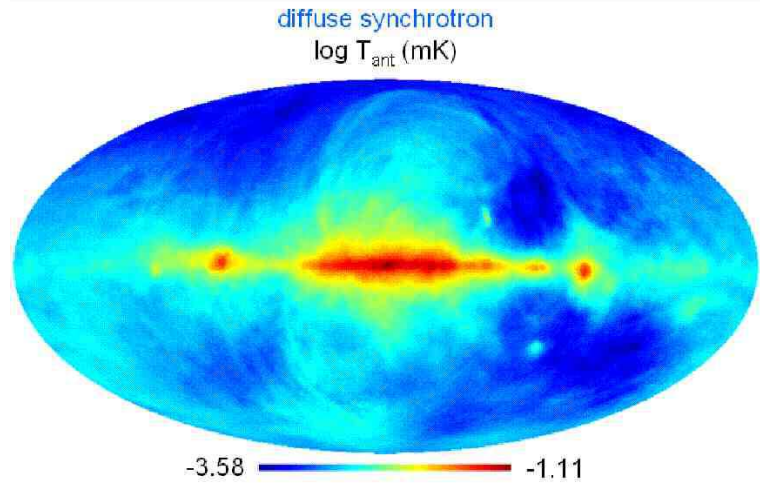}
   \caption{Map of the Galactic diffuse synchrotron emission 
adopted in this study (see also the text).}
%              \label{FigGam}%
    \end{figure*}
   \begin{figure*}[!h]
   \centering
   \includegraphics[width=10cm]{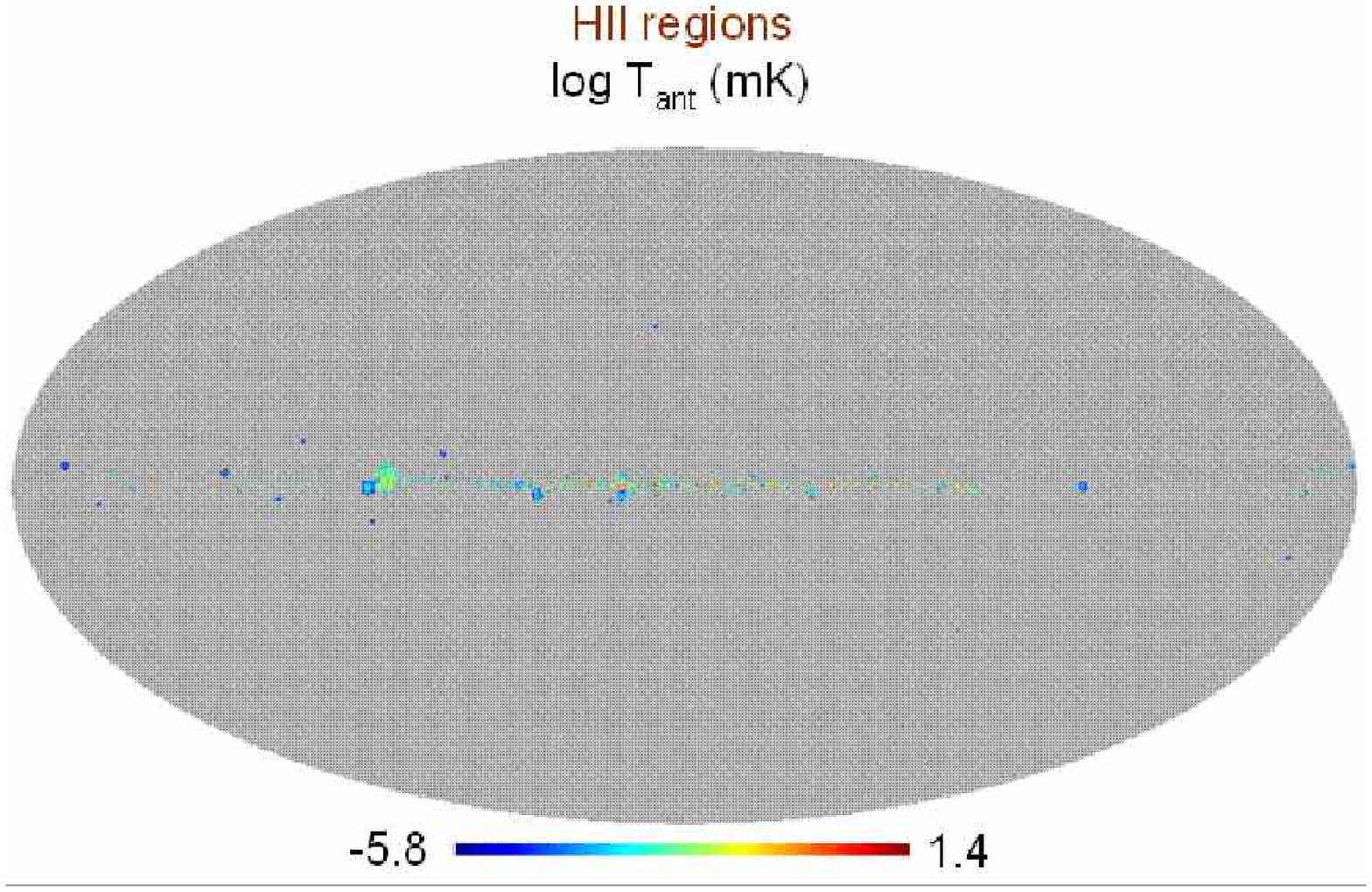}
   \caption{Map of the free-free emission from compact Galactic HII regions
adopted in this study (see also the text).}
%              \label{FigGam}%
    \end{figure*}

Localized free-free emission dominates the signal 
on wide areas in the Galactic plane
at least over the frequency range from 30 to 100~GHz. 
The Synthetic Catalog at 2.7~GHz produced by Paladini et al. (2003) 
provides a rich information on compact Galactic HII regions 
that has been used in this work to generate a map of
free-free emission from these sources (see Fig.~3).
A spectral index $\alpha =-0.1$ in flux 
($-2.1$ in antenna temperature), as in the case of thermal bremsstrahlung
emission in a thin plasma, has been adopted here 
to extrapolate the signal from 2.7~GHz to the considered 
{\sc Planck} frequency channels.
With respect to the simulated map reported by Paladini et al. (2003),
we have implemented here a code which simulates
the contribution of each source in the Synthetic Catalog to each map pixel 
without applying at this step the convolution with the beam, 
since the convolution with the intermediate and far antenna pattern 
has been subsequently applied as described in the first part of this section.

All maps have been projected into the HEALPix 
scheme~\footnote{http://www.eso.org/science/healpix/}
({\it Hierarchical Equal Area and IsoLatitude Pixelization
of the Sphere}) by G\`orski et al. (1999).

While this work was nearly completed, the 1~year 
data products~\footnote{http://lambda.gsfc.nasa.gov} 
from the WMAP satellite have become available.
In a few representative cases, we have repeated the straylight analysis
by adopting the WMAP frequency maps at 33 and 94~GHz, including
in the straylight evaluation also the minor contributions to 
straylight signal from CMB and extragalactic source 
fluctuations~\footnote{Note that these WMAP maps do not include the monopole, 
while our Galactic component maps include the corresponding monopoles. The 
comparison between the average straylight signals will be therefore 
only indicative. Clearly, anisotropy experiments are not sensitive to the monopole,
directly subtracted in the data.}.
Note that WMAP maps are, of course, convolved with the 
corresponding beam patterns and include the effect of main beam
distortions and straylight contamination as well as the instrumental 
noise and the effect of other sistematics not subtracted in the 
data analysis; on the other hand,
these effects are significantly smaller than the signal 
(Hinshaw et al. 2003a) and, 
since this analysis is mainly contributed by signal variations 
on degree or larger angular scales, they can be neglected.

\section{Simulation results}

The main output of our simulation code are the time ordered data 
(TOD) of the signals entering the intermediate and far pattern.

In the TOD of each scan circle, two prominent maxima typically appear. 
These are related, for the intermediate pattern, to the two crossings
of the Galactic plane of the telescope field of view and to the
crossings of the Galactic plane of the main spillover in the
case of the far pattern.
As already recognized by Burigana et al. (2001), these maxima 
are only slighlty shifted with 
respect to the maxima of the signal entering the main beam
in the case of the intermediate pattern and shifted of about
$90^{\circ}$ in the case of the far pattern, as direct
consequence of the pattern shape.
% (see Figs.~2 and 3).

The typical signal level is determined by the sky signal 
and the fraction, 
$f_\%=100\int_{\underline{\Omega}} J d\Omega / \int_{4\pi} J d\Omega$, 
of integrated antenna response in the considered
antenna pattern region, $\underline{\Omega}$, 
reported in Table~3 of Paper~I.
The ratio between the fraction of the integrated antenna response 
in the far pattern and the one in the intermediate pattern 
provides only a rough upper limit to the ratio of straylight peak-to-peak 
signal in these two pattern regions. In fact, 
while an extended very bright Galactic region could quite easily 
fill the (relatively small) solid angle subtended
by the intermediate pattern,   
the whole (quite large) solid angle subtented 
by the main spillover and by the other relevant far pattern features 
can not be easily filled by signals coming all simultaneously from 
very bright Galactic regions.
Similarly, the fractional difference between the integrated antenna response
in the far pattern computed by including or not the third order optical
interaction provides an upper limit (less than about 10\%) 
to the fractional underestimation of the GSC when the third order 
optical interaction is neglected. More accurate estimates
require numerical simulations.

In Figs.~4 and 5 we report the TOD corresponding to the straylight
signal respectively from the far and intermediate pattern 
for the three Galactic components described in Sect.~2.2 in the case
of the beam LFI9 9B computed by including the first, second and third order 
optical interactions. 

   \begin{figure*}[t]
   \centering
   \begin{tabular}{ccc}
   \includegraphics[width=5cm]{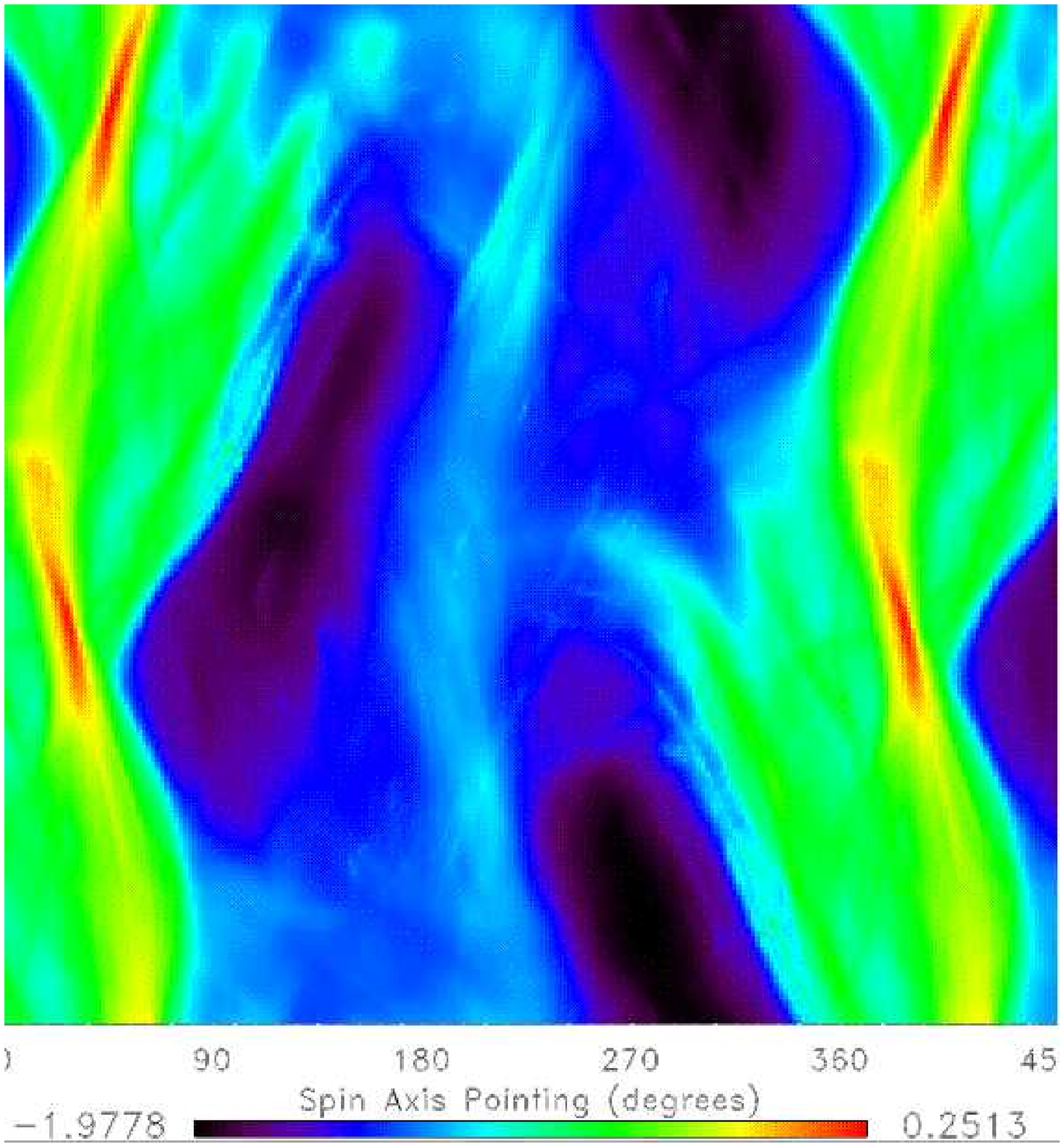}&
   \includegraphics[width=5cm]{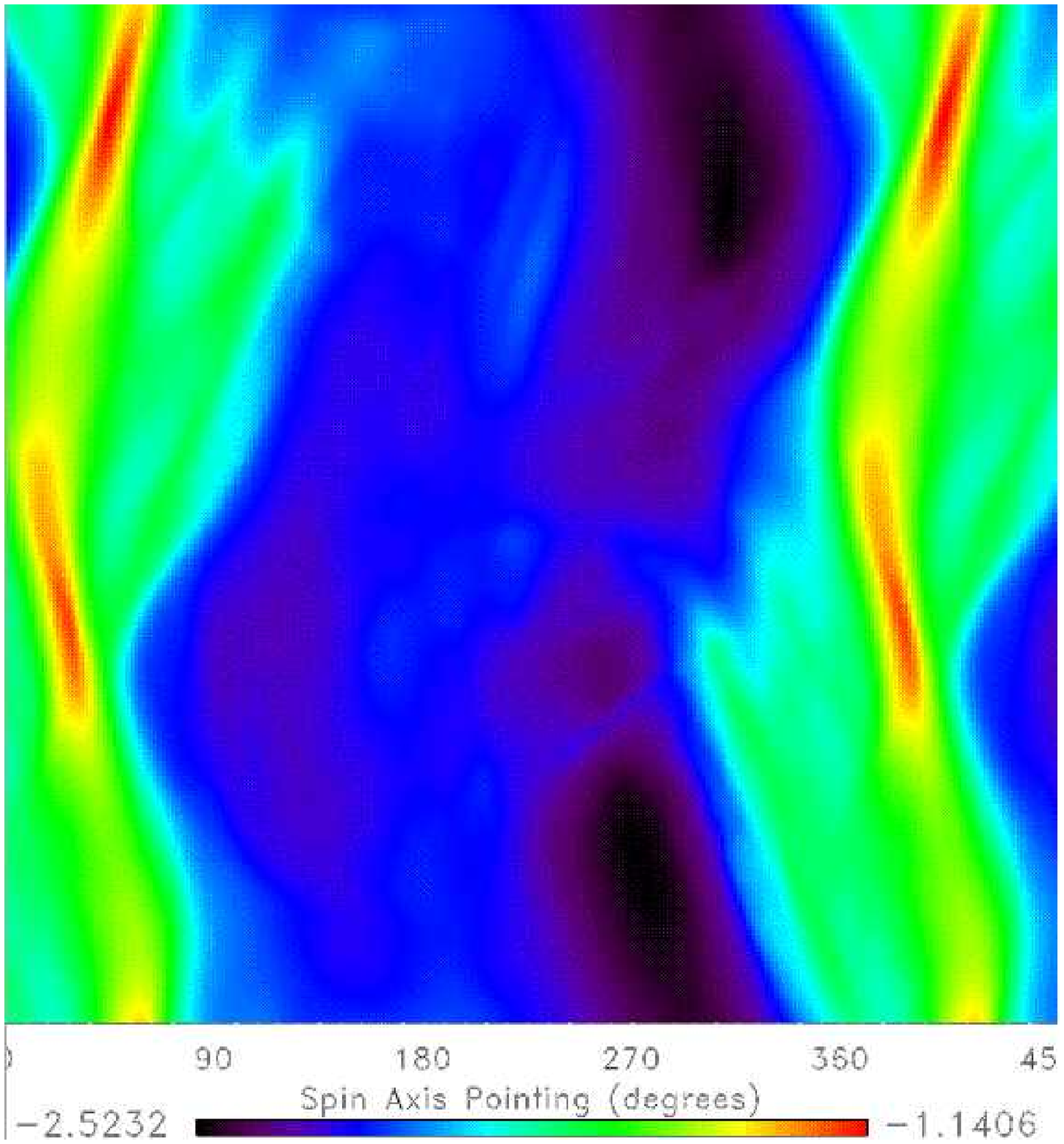}&
   \includegraphics[width=5cm]{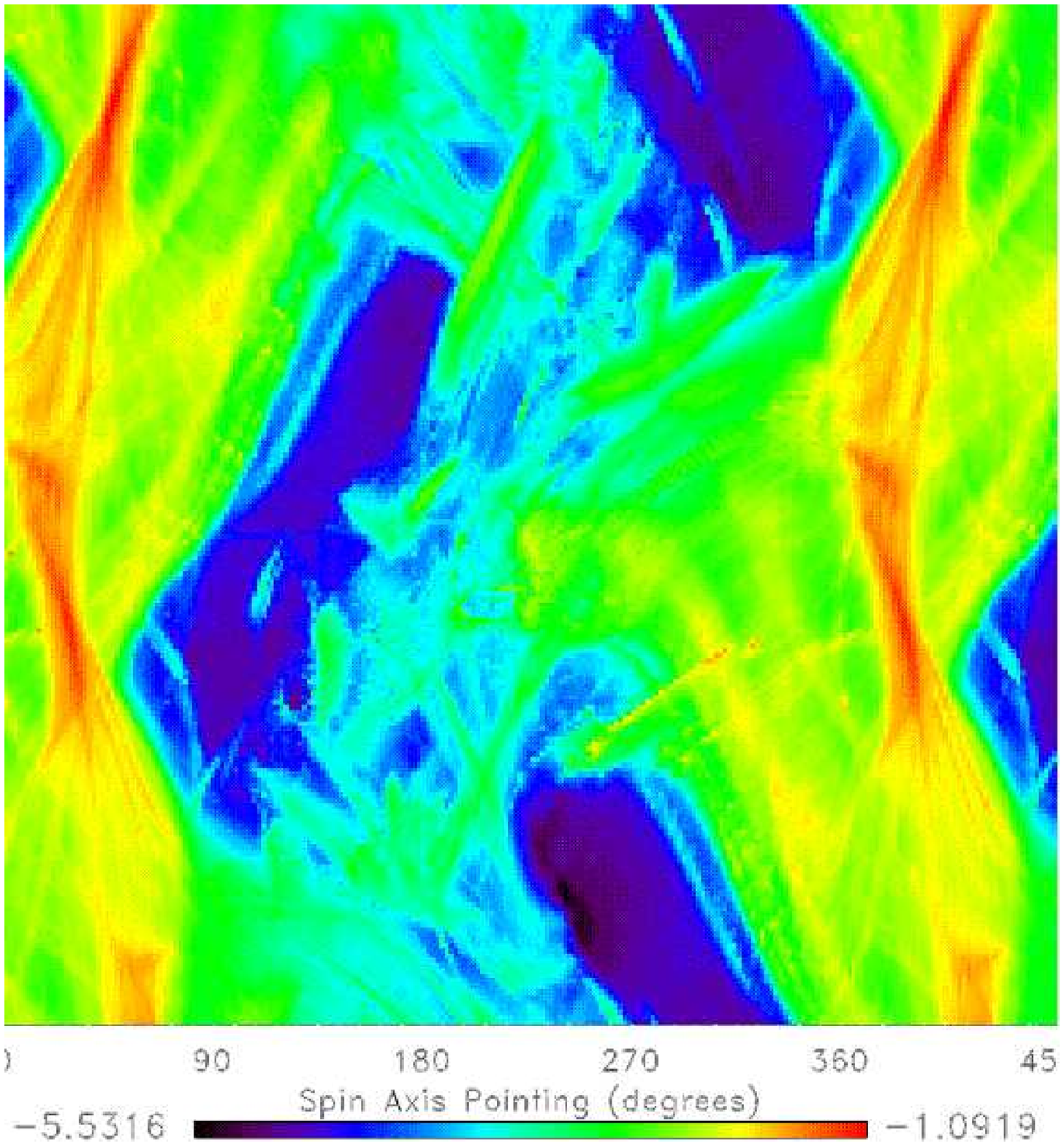}
   \end{tabular}
   \caption{Synthetic view of the data stream in terms
of decimal logarithm of the antenna temperature in $\mu$K from all scan
circles for the different Galactic components. The ecliptic coordinates 
properly refer here to the latitude of the telescope LOS
(for graphic purposes, in this plot the range 
between $-85^{\circ}$ and $-255^{\circ}$ 
refer to the second half of each scan circle)
and to the its longitude shift with respect to its initial direction,
or equivalently to the shift of the spin axis pointing direction.
We report here the straylight signal in the far sidelobes computed
by including the first, second and third order optical interactions
for the beam LFI9 9B (see also the text).}
%              \label{FigGam}%
    \end{figure*}
   \begin{figure*}[!h]
   \centering
   \begin{tabular}{ccc}
   \includegraphics[width=5cm]{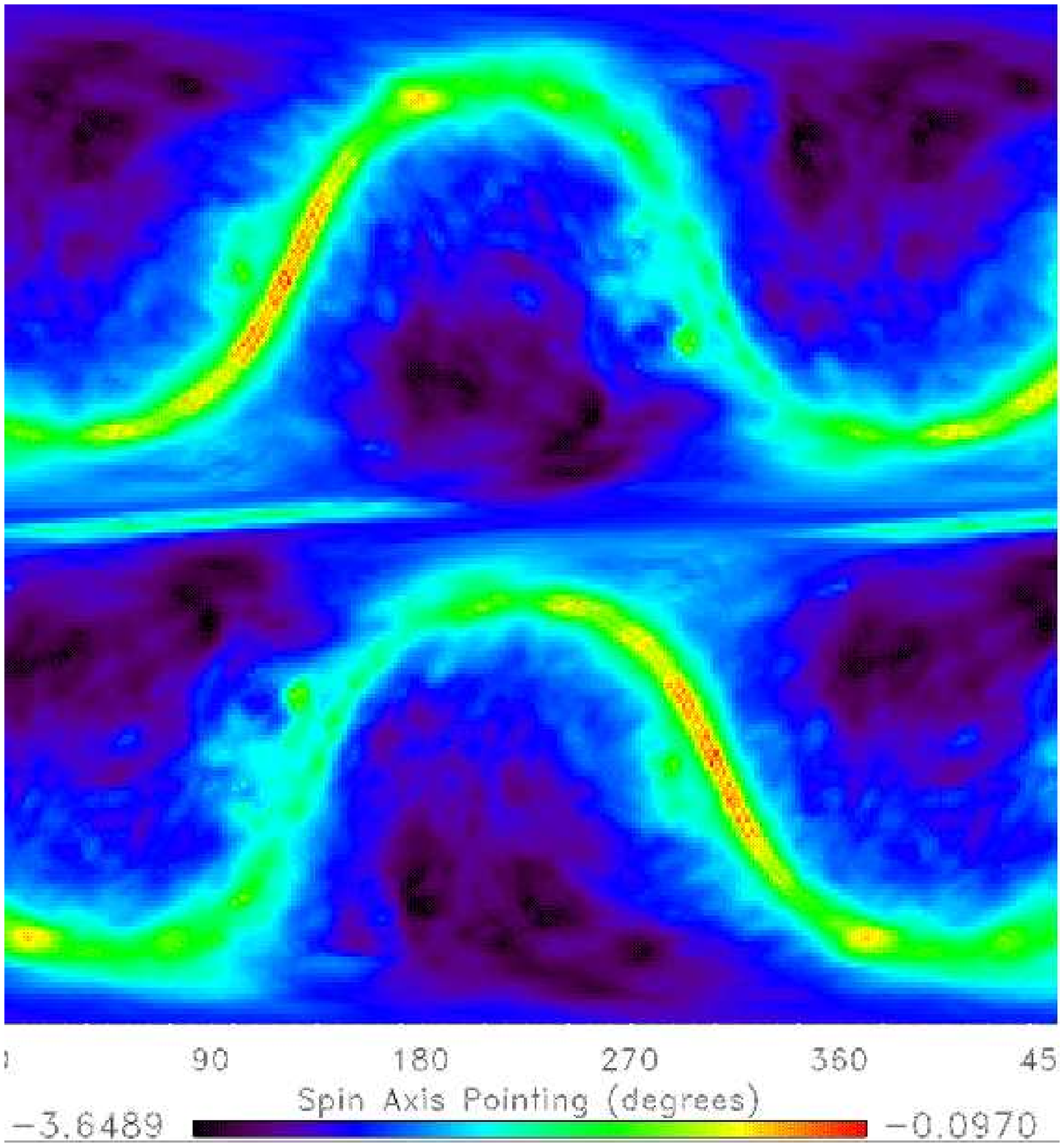}&
   \includegraphics[width=5cm]{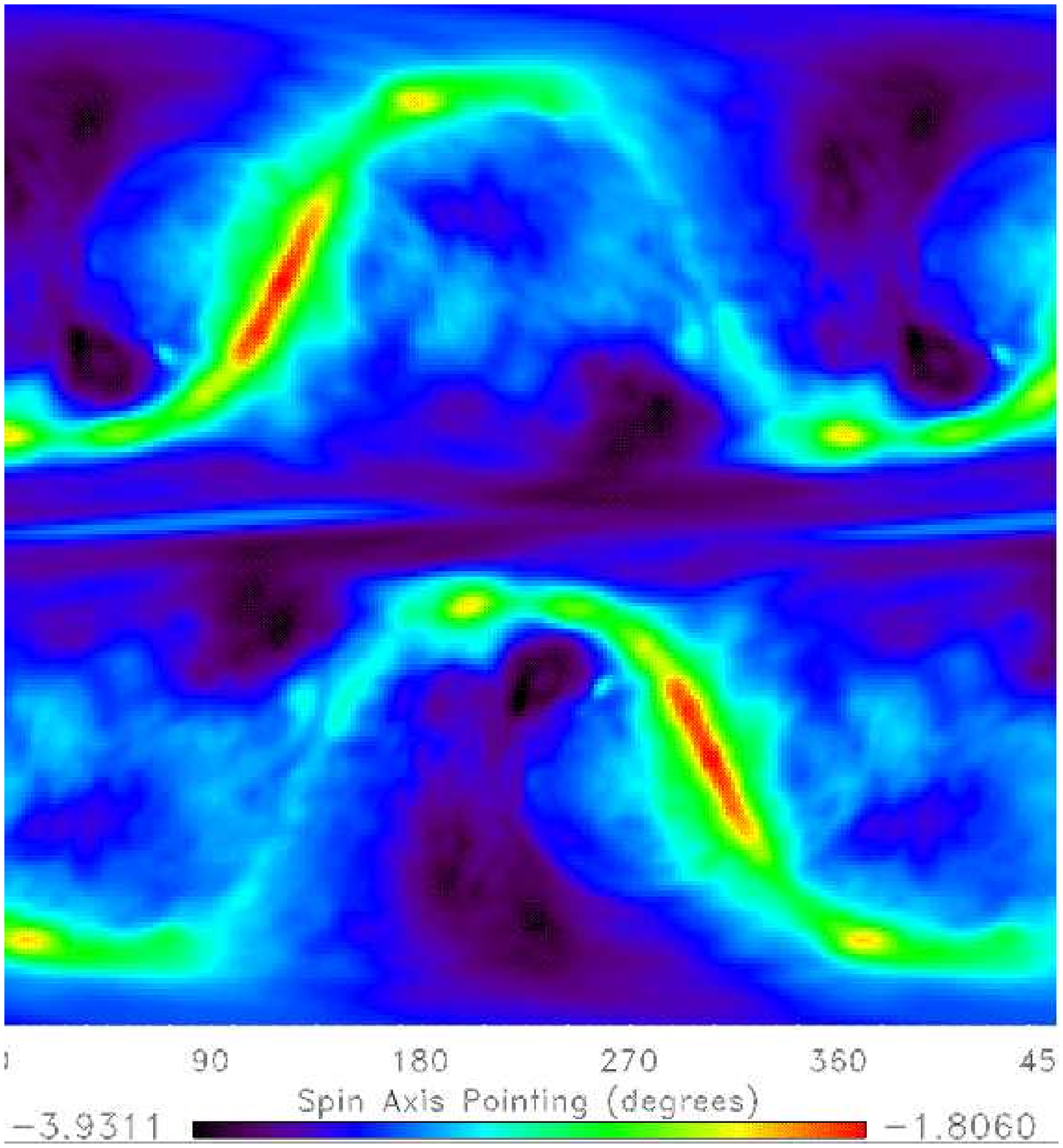}&
   \includegraphics[width=5cm]{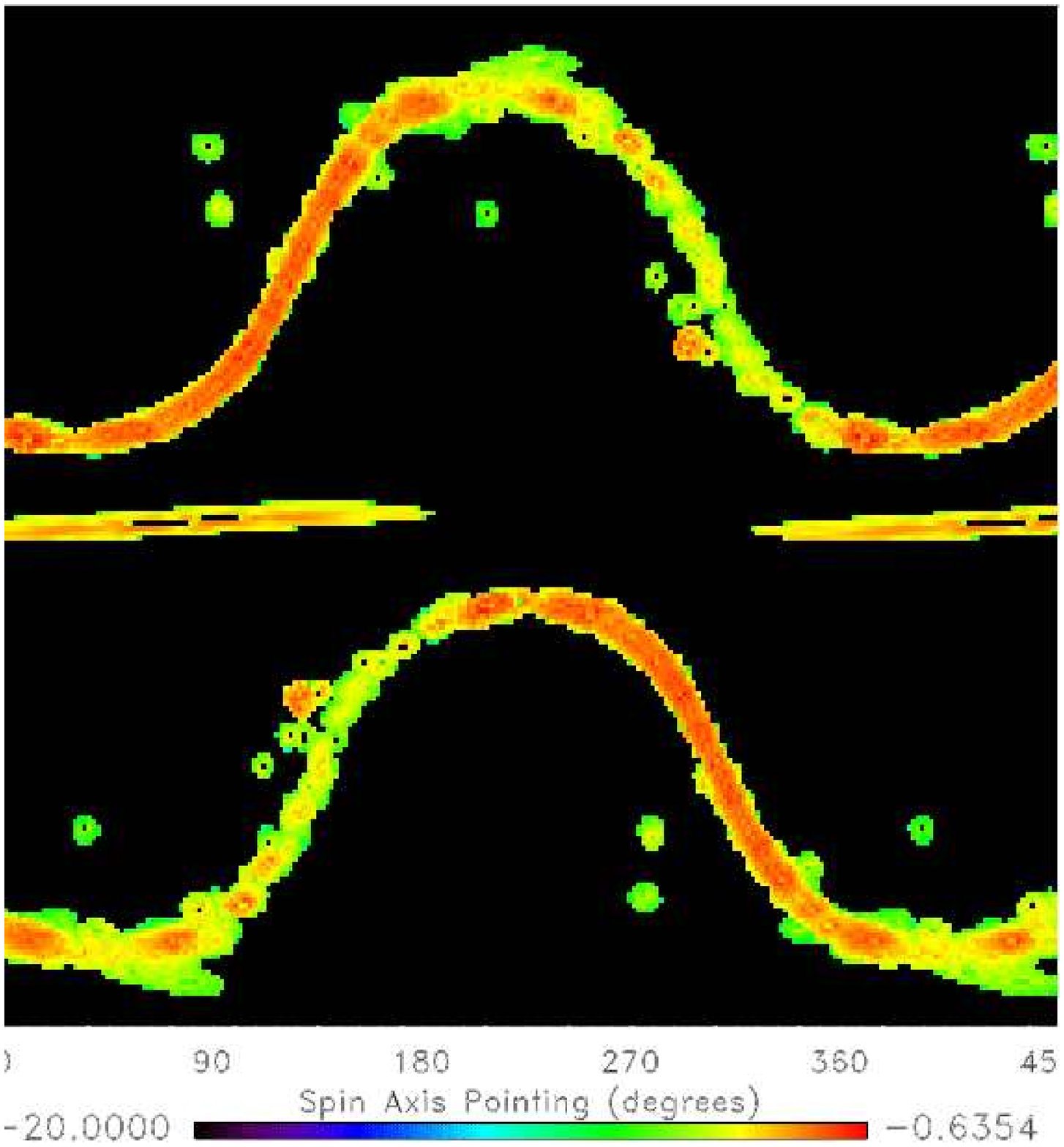}
   \end{tabular}
   \caption{The same as in Fig.~4, but for the signal from the intermediate pattern.
In this case, Galactic HII regions produce a quite localized straylight contamination,
null far from these sources and arbitrarily set to $10^{-20}\mu$K
for graphic purposes (as in Fig.~4, we report here the
decimal logarithm of the antenna temperature in $\mu$K; 
see also the text).}
%              \label{FigGam}%
    \end{figure*}

The TOD corresponding to the difference between the straylight signal 
obtained by including or neglecting the third order optical interaction 
are reported in  Figs.~6 and 7.
Clearly, the difference is within $\sim 5$~\% of the straylight
signal, i.e. smaller than the above upper limit derived on the basis of simple 
optical considerations by about a factor two. 
Therefore, considering only the first and second order optical interactions
does not introduce relevant loss of information in the optimization
analysis of the optical 
design~\footnote{On the contrary, as obvious, 
the best possible knowledge of the antenna
pattern, including all possible effects, from those due dust and molecular
contamination on mirror surfaces to those related to the mirror roughness 
and temperature behaviour, should to be taken into account in 
the final data analysis.}.

   \begin{figure*}[t]
   \centering
   \begin{tabular}{ccc}
   \includegraphics[width=5cm]{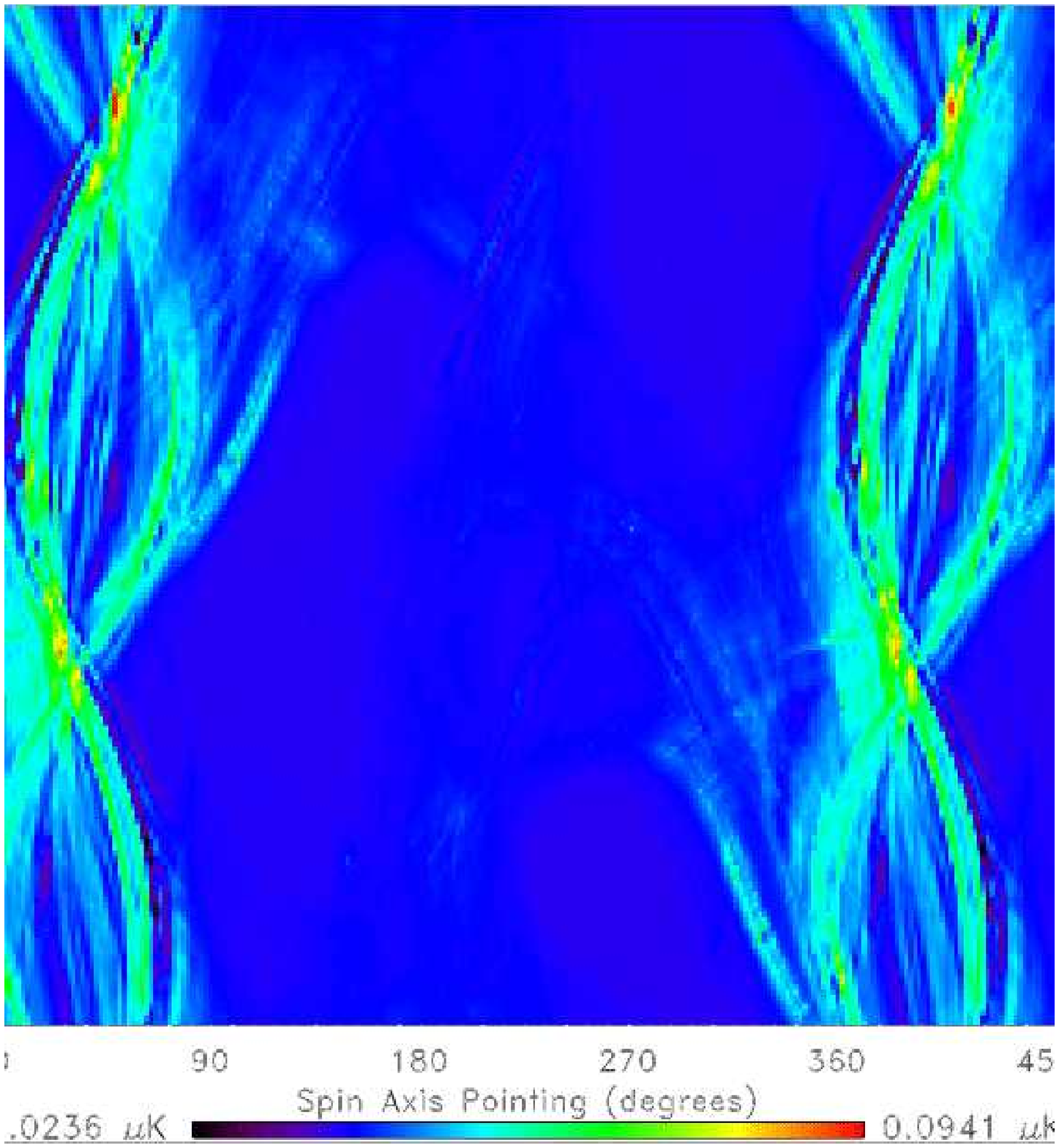}&
   \includegraphics[width=5cm]{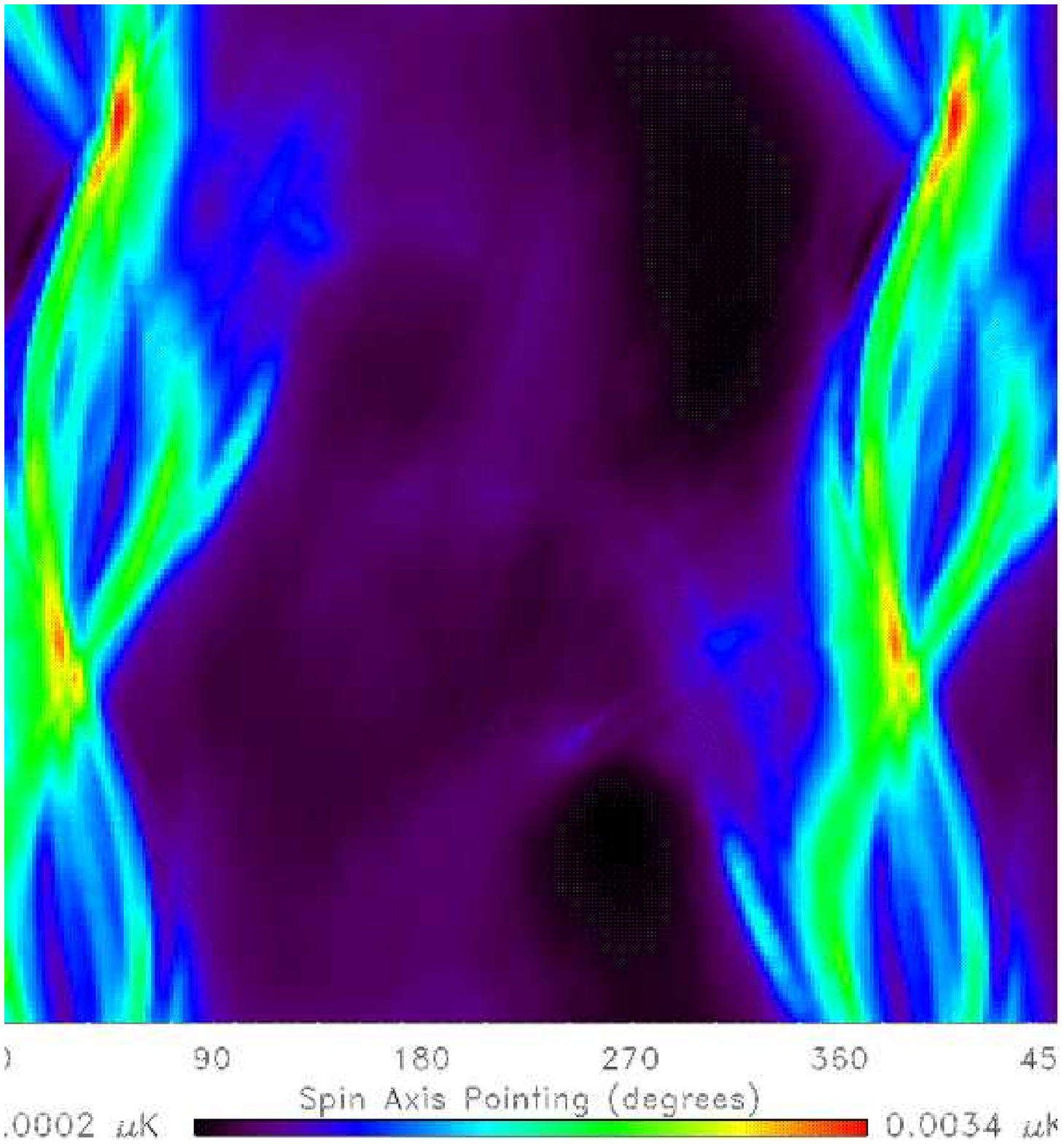}&
   \includegraphics[width=5cm]{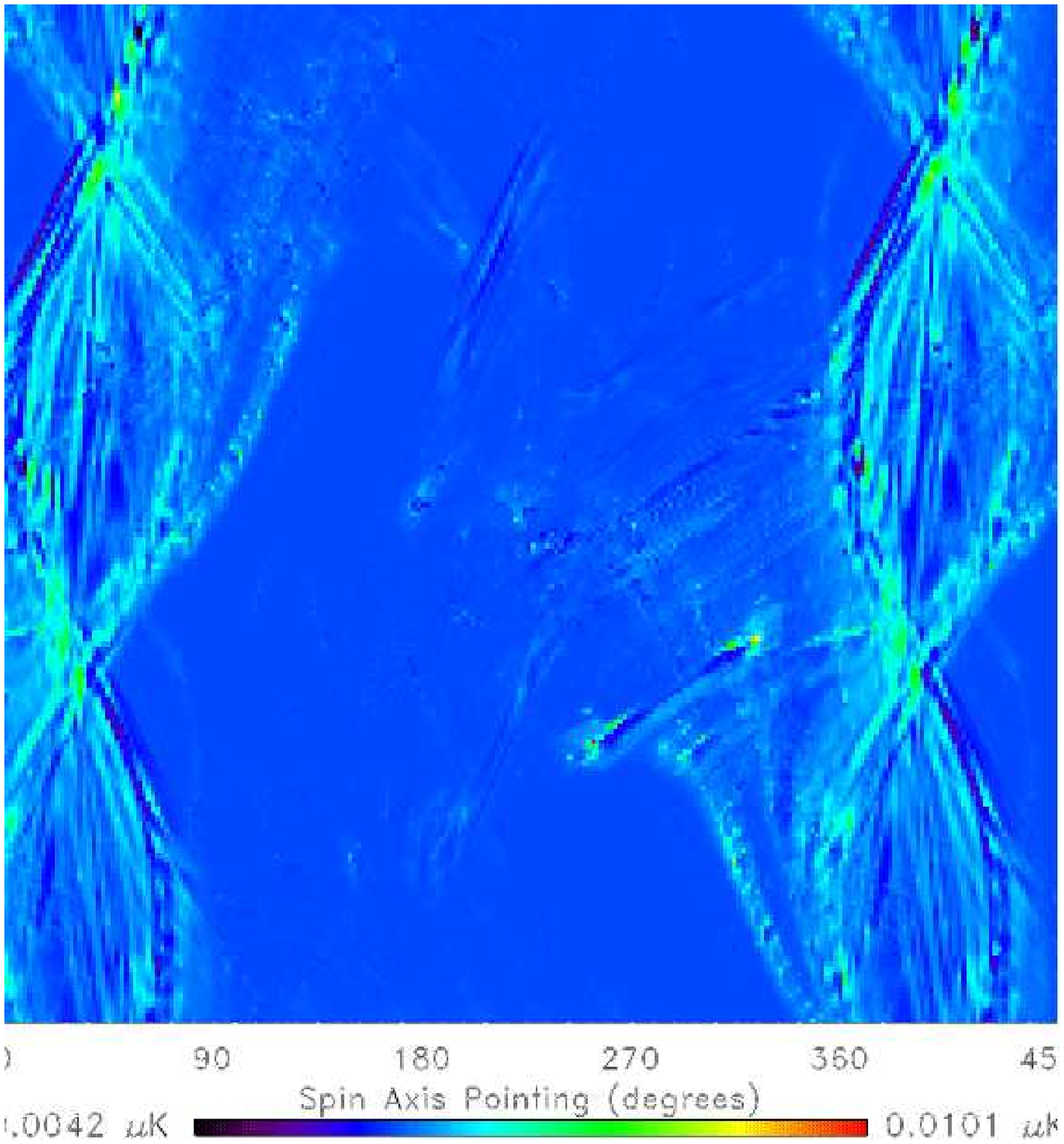}
   \end{tabular}
   \caption{The same as in Fig.~4, but referring to  
the difference between the straylight signals computed by 
including or not the third order optical interaction 
(the antenna temperature in $\mu$K in linear scale
is here reported; see also the text).}
%              \label{FigGam}%
    \end{figure*}
   \begin{figure*}[!h]
   \centering
   \begin{tabular}{ccc}
   \includegraphics[width=5cm]{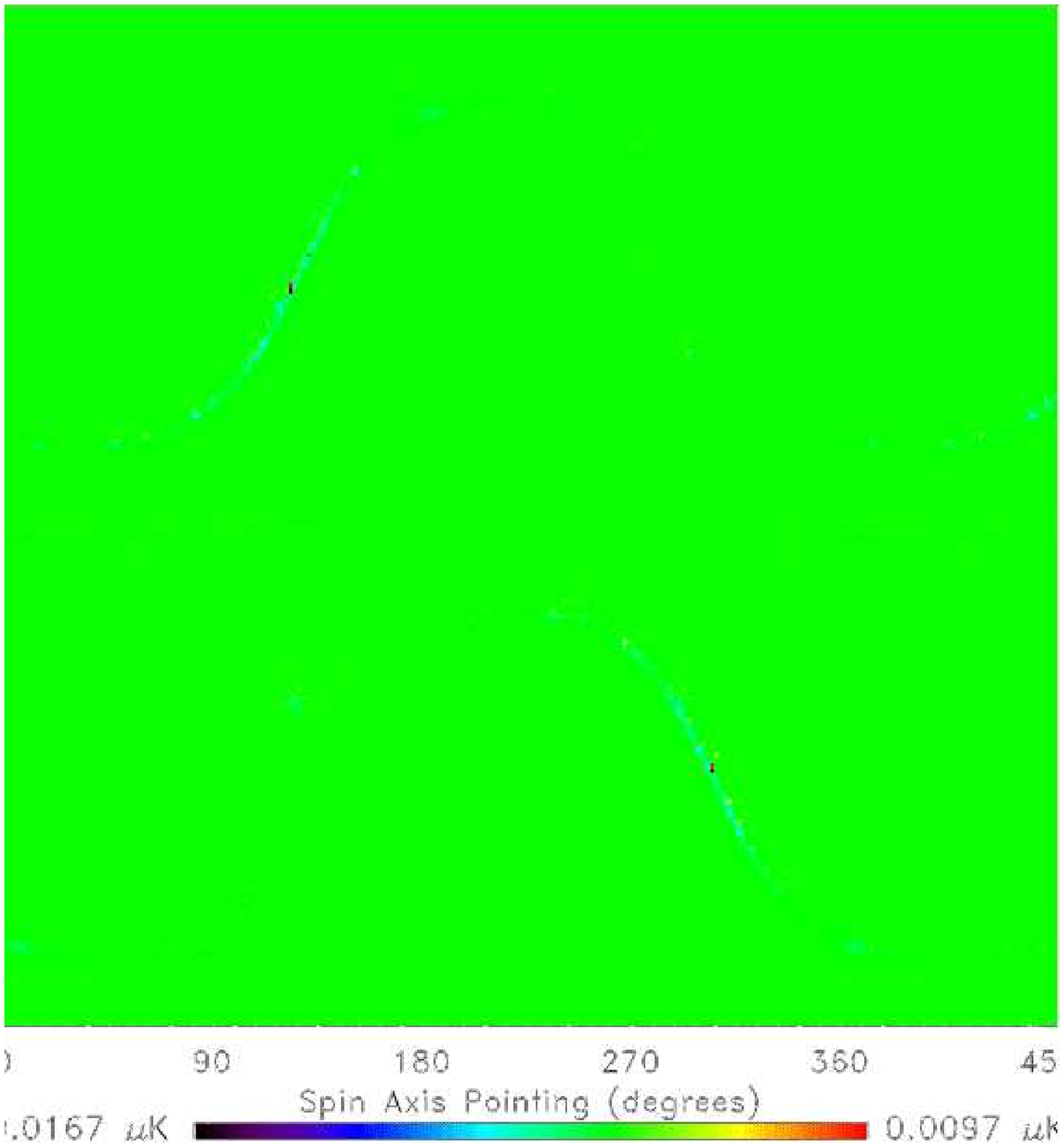}&
   \includegraphics[width=5cm]{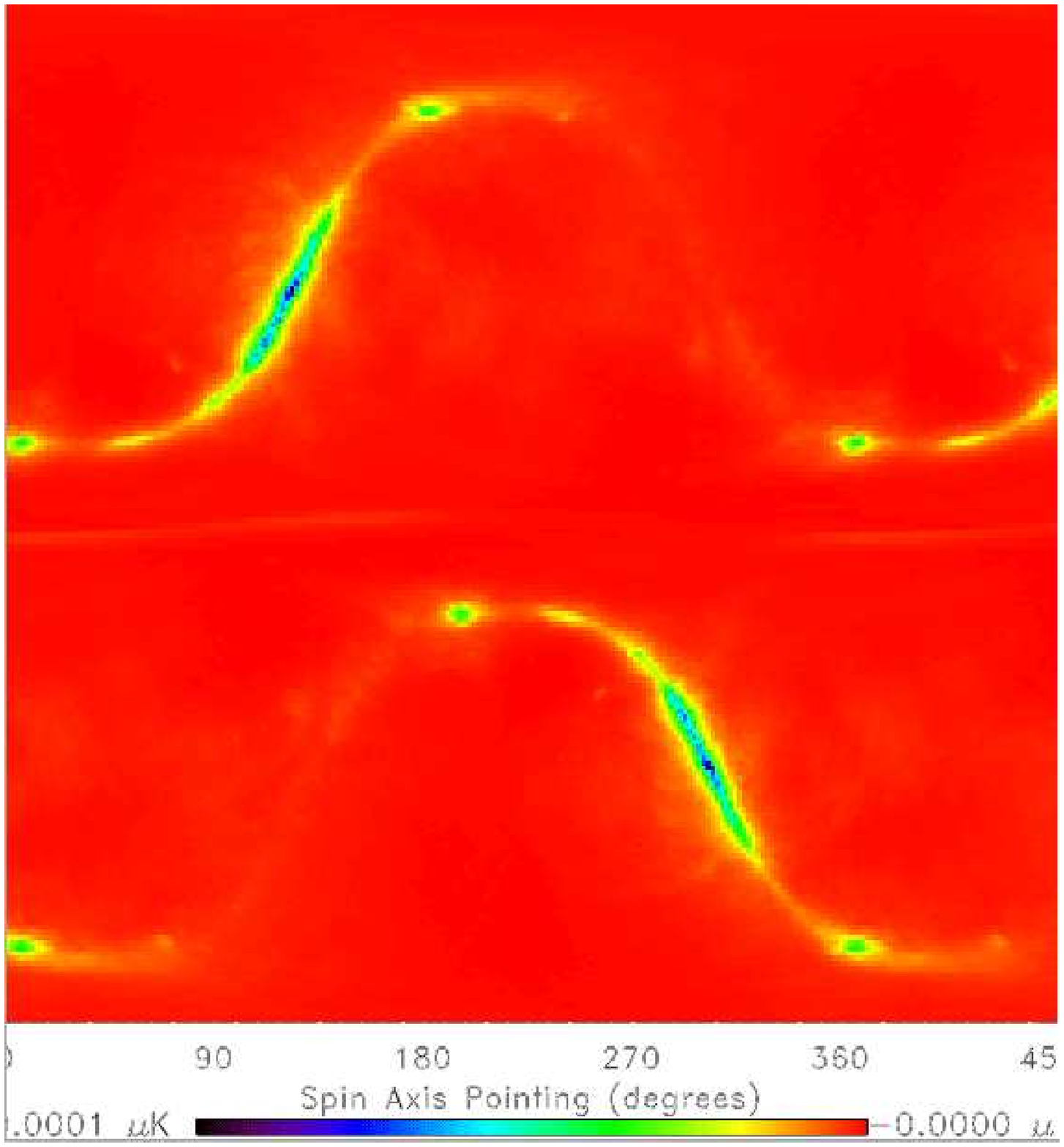}&
   \includegraphics[width=5cm]{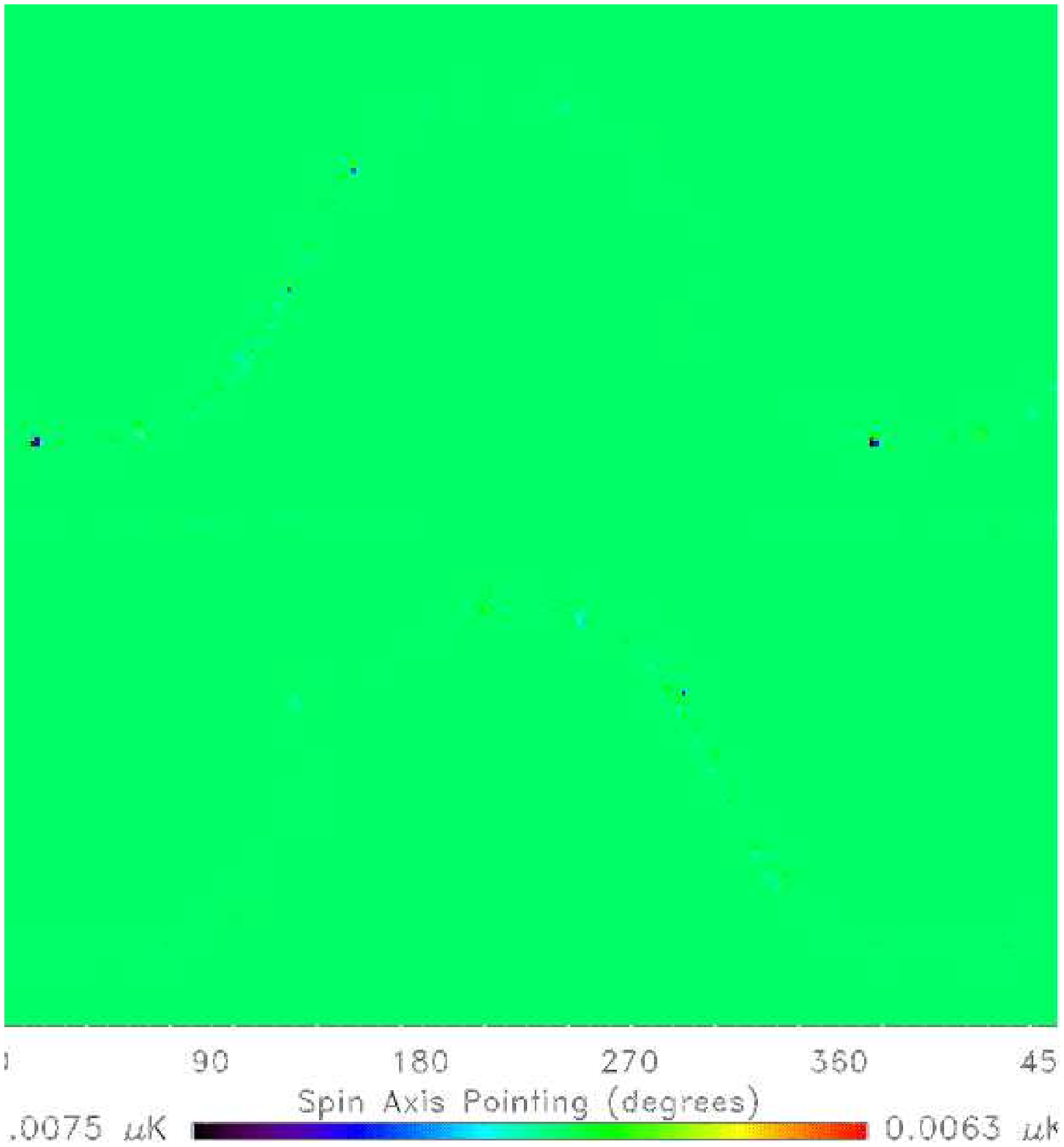}
   \end{tabular}
   \caption{The same as in Fig.~6, but for the signal from the intermediate pattern
(see also the text).}
%              \label{FigGam}%
    \end{figure*}
 
By comparing Figs.~1, 2, and 3 with Figs.~4 and 5, note how the  
different angular distribution of the three considered 
Galactic components reflects in the pattern of the straylight
signal TOD, more or less ``diffuse'' according to the considered component. 

Similar results are found for all the considered beam patterns, but with a
peak-to-peak and {\sc rms} straylight signal significantly dependent on the 
adopted pattern shape. 
Instead of reporting the full simulation results, 
a concise comparison between the 
results obtained for the whole set of 
optical configurations is reported in the 
next section.

\section{Comparison between different antenna patterns and 
foreground components}

The statistical moments of the straylight signal TOD 
and its peak-to-peak value are reported in Tables~1--7 
for our whole set of optical configurations at 100~GHz
[in the tables, the data referring to the global straylight effect from
intermediate pattern plus far sidelobes (I + F) are
derived including also the contribution from the third order optical interactions, 
when available]. Table~8 summarizes the basic information contained 
in the above tables as functions of the edge taper.

As evident from the tables, the contamination from the far sidelobes is 
much more relevant than that from the intermediate pattern 
when the {\sc rms} of the TOD is considered; by looking at the peak-to-peak 
of the straylight signal, 
the contamination from the far sidelobes is larger by a factor of few units 
or comparable to that from the intermediate pattern.
The larger impact of the far sidelobes with respect to the intermediate pattern
is particularly remarkable for the case of the diffuse 
Galactic components, while it is less evident in the case the 
map of free-free emission from Galactic HII regions.
We find in fact a quite general behaviour: more diffuse the component and more 
relevant the straylight contamination from far sidelobes with respect
to the intermediate pattern. 

On the contrary, the skewness and kurtosis indices are larger for the
straylight contamination from the intermediate pattern because 
they are more sensitive to localized features. 
As expected, these indices 
are larger for the free-free emission from Galactic HII regions
than for the more diffuse components.

% //////////// LFI9 PG25=9A ////////////
\begin{table}
\centering
\caption{Statistical moments and peak-to-peak of the simulated TOD 
(antenna temperature expressed in $\mu$K) from intermediate
and far sidelobes of the simulated beam pattern LFI9 9A at 100~GHz for different 
Galactic components. The row labelled with ``I123'' (``I12'') 
refers to the signal in the intermediate
beam computed taking into account the first, second and third (first and second)
order optical interactions.
The row labelled with ``F123'' (``F12'') refers to to the signal in the far sidelobes
computed taking into account the first, second and third (first and second) order
optical interactions (see also the text).}
\begin{tabular}{c c c c c c c}
\hline
\hline
\multicolumn{7}{c}{{\sc LFI9 9A}}  \\
{\sc beam}	& &	&	&	&	{\sc skewness}	&	{\sc kurtosis} \\
{\sc region}	&	{\sc average} &	{\sc variance} & {\sc rms} &	{\sc peak-to-peak} & {\sc index}	&	{\sc index} \\
\hline
\hline
\multicolumn{7}{l} {\sc dust + diffuse free--free emission}  \\
\hline
I12&  2.27$\times 10^{-3}$&  5.81$\times 10^{-5}$&  7.63$\times 10^{-3}$&  2.61$\times 10^{-1}$&  9.45$\times 10^{+0}$&  1.33$\times 10^{+2}$\\	  
F123&  6.71$\times 10^{-2}$&  4.77$\times 10^{-3}$&  6.90$\times 10^{-2}$&  5.88$\times 10^{-1}$&  1.88$\times 10^{+0}$&  4.92$\times 10^{+0}$\\
F12&  6.40$\times 10^{-2}$&  4.49$\times 10^{-3}$&  6.70$\times 10^{-2}$&  5.68$\times 10^{-1}$&  1.89$\times 10^{+0}$&  4.90$\times 10^{+0}$\\
I + F& 6.93$\times 10^{-2}$&  4.83$\times 10^{-3}$&  6.95$\times 10^{-2}$&  5.88$\times 10^{-1}$&  1.81$\times 10^{+0}$&  4.59$\times 10^{+0}$\\
\hline
\multicolumn{7}{l}{\sc diffuse synchrotron emission }  \\
\hline
I12&  2.10$\times 10^{-4}$&  1.05$\times 10^{-7}$&  3.24$\times 10^{-4}$&  4.66$\times 10^{-3}$&  6.59$\times 10^{+0}$&  5.64$\times 10^{+1}$\\
F123&  5.87$\times 10^{-3}$&  1.65$\times 10^{-5}$&  4.07$\times 10^{-3}$&  2.32$\times 10^{-2}$&  1.27$\times 10^{+0}$&  1.11$\times 10^{+0}$\\
F12&  5.60$\times 10^{-3}$&  1.55$\times 10^{-5}$&  3.94$\times 10^{-3}$&  2.24$\times 10^{-2}$&  1.29$\times 10^{+0}$&  1.13$\times 10^{+0}$\\
I + F& 6.08$\times 10^{-3}$&  1.67$\times 10^{-5}$&  4.09$\times 10^{-3}$&  2.31$\times 10^{-2}$&  1.22$\times 10^{+0}$&  9.44$\times 10^{-1}$\\
\hline
\multicolumn{7}{l}{\sc HII regions }  \\
\hline
I12&  7.70$\times 10^{-5}$&  7.45$\times 10^{-7}$&  8.63$\times 10^{-4}$&  8.05$\times 10^{-2}$&  4.14$\times 10^{+1}$&  2.67$\times 10^{+3}$\\
F123&  2.13$\times 10^{-3}$&  7.14$\times 10^{-6}$&  2.67$\times 10^{-3}$&  2.67$\times 10^{-2}$&  1.93$\times 10^{+0}$&  5.87$\times 10^{+0}$\\
F12&  2.02$\times 10^{-3}$&  6.69$\times 10^{-6}$&  2.59$\times 10^{-3}$&  2.61$\times 10^{-2}$&  1.94$\times 10^{+0}$&  5.82$\times 10^{+0}$\\
I + F& 2.20$\times 10^{-3}$&  7.99$\times 10^{-6}$&  2.83$\times 10^{-3}$&  8.67$\times 10^{-2}$&  2.97$\times 10^{+0}$&  3.27$\times 10^{+1}$\\
\hline
\multicolumn{7}{l}{\sc sum of the above components}  \\
\hline
I12&  2.55$\times 10^{-3}$&  7.02$\times 10^{-5}$&  8.38$\times 10^{-3}$&  2.94$\times 10^{-1}$&  9.51$\times 10^{+0}$&  1.37$\times 10^{+2}$\\
F123&  7.51$\times 10^{-2}$&  5.71$\times 10^{-3}$&  7.56$\times 10^{-2}$&  6.38$\times 10^{-1}$&  1.84$\times 10^{+0}$&  4.66$\times 10^{+0}$\\
F12&  7.16$\times 10^{-2}$&  5.38$\times 10^{-3}$&  7.33$\times 10^{-2}$&  6.16$\times 10^{-1}$&  1.85$\times 10^{+0}$&  4.64$\times 10^{+0}$\\
I + F& 7.76$\times 10^{-2}$&  5.79$\times 10^{-3}$&  7.61$\times 10^{-2}$&  6.38$\times 10^{-1}$&  1.78$\times 10^{+0}$&  4.34$\times 10^{+0}$\\
\hline
\hline
\end{tabular}
\end{table}

% //////////// LFI9 PG27=9B ////////////
\begin{table}
\centering
\caption{The same as in Table~1, but for the indicated simulated beam pattern.
Note how the straylight in the intermediate pattern 
I123 the GSC is just smaller when the third order optical interaction is included 
(compare I123 with I12).
This is due to a small decrease ($\simeq 0.4$~\%) of the integrated antenna pattern
response in that pattern region when the third order optical interaction is taken into account  
because of the combination in amplitude and phase of the various contributions, 
producing an averall antenna response
-- different from a simple sum of the powers of the various contributions -- that typically
increases with the number of considered contributions, 
as occurs here in the case of far sidelobes, but not necessarily does.
For the straylight in the intermediate pattern we report also the result based on 
the WMAP map at 94~GHz, including all components.}
\begin{tabular}{c c c c c c c}
\hline
\hline
\multicolumn{7}{c}{{\sc LFI9 9B}}  \\
{\sc beam}	& &	&	&	&	{\sc skewness}	&	{\sc kurtosis} \\
{\sc region}	&	{\sc average} &	{\sc variance} & {\sc rms} &	{\sc peak-to-peak} & {\sc index}	&	{\sc index} \\
\hline
\hline
\multicolumn{7}{l} {\sc dust + diffuse free--free emission}  \\
\hline
I123&  7.54$\times 10^{-3}$&  6.38$\times 10^{-4}$&  2.53$\times 10^{-2}$&  8.00$\times 10^{-1}$&  9.24$\times 10^{+0}$&  1.23$\times 10^{+2}$\\
I12&  7.57$\times 10^{-3}$&  6.41$\times 10^{-4}$&  2.53$\times 10^{-2}$&  7.90$\times 10^{-1}$&  9.18$\times 10^{+0}$&  1.21$\times 10^{+2}$\\
F123&  1.45$\times 10^{-1}$&  3.31$\times 10^{-2}$&  1.82$\times 10^{-1}$&  1.77$\times 10^{+0}$&  2.88$\times 10^{+0}$&  1.19$\times 10^{+1}$\\
F12&  1.35$\times 10^{-1}$&  3.03$\times 10^{-2}$&  1.74$\times 10^{-1}$&  1.73$\times 10^{+0}$&  2.96$\times 10^{+0}$&  1.26$\times 10^{+1}$\\
I + F& 1.53$\times 10^{-1}$&  3.35$\times 10^{-2}$&  1.83$\times 10^{-1}$&  1.77$\times 10^{+0}$&  2.77$\times 10^{+0}$&  1.12$\times 10^{+1}$\\
\hline
\multicolumn{7}{l}{\sc diffuse synchrotron emission }  \\
\hline
I123&  6.96$\times 10^{-4}$&  1.16$\times 10^{-6}$&  1.08$\times 10^{-3}$&  1.55$\times 10^{-2}$&  6.58$\times 10^{+0}$&  5.63$\times 10^{+1}$\\
I12&  6.99$\times 10^{-4}$&  1.17$\times 10^{-6}$&  1.08$\times 10^{-3}$&  1.55$\times 10^{-2}$&  6.57$\times 10^{+0}$&  5.61$\times 10^{+1}$\\
F123&  1.27$\times 10^{-2}$&  1.06$\times 10^{-4}$&  1.03$\times 10^{-2}$&  6.93$\times 10^{-2}$&  1.89$\times 10^{+0}$&  3.97$\times 10^{+0}$\\
F12&  1.18$\times 10^{-2}$&  9.63$\times 10^{-5}$&  9.81$\times 10^{-3}$&  6.71$\times 10^{-2}$&  1.95$\times 10^{+0}$&  4.24$\times 10^{+0}$\\
I + F& 1.34$\times 10^{-2}$&  1.07$\times 10^{-4}$&  1.03$\times 10^{-2}$&  6.91$\times 10^{-2}$&  1.81$\times 10^{+0}$&  3.67$\times 10^{+0}$\\
\hline
\multicolumn{7}{l}{\sc HII regions }  \\
\hline
I123&  2.58$\times 10^{-4}$&  7.31$\times 10^{-6}$&  2.70$\times 10^{-3}$&  2.32$\times 10^{-1}$&  3.57$\times 10^{+1}$&  2.03$\times 10^{+3}$\\
I12&  2.59$\times 10^{-4}$&  7.31$\times 10^{-6}$&  2.70$\times 10^{-3}$&  2.32$\times 10^{-1}$&  3.55$\times 10^{+1}$&  2.02$\times 10^{+3}$\\
F123&  4.45$\times 10^{-3}$&  4.70$\times 10^{-5}$&  6.86$\times 10^{-3}$&  8.09$\times 10^{-2}$&  3.18$\times 10^{+0}$&  1.60$\times 10^{+1}$\\
F12&  4.13$\times 10^{-3}$&  4.27$\times 10^{-5}$&  6.54$\times 10^{-3}$&  7.64$\times 10^{-2}$&  3.27$\times 10^{+0}$&  1.68$\times 10^{+1}$\\
I + F& 4.71$\times 10^{-3}$&  5.45$\times 10^{-5}$&  7.38$\times 10^{-3}$&  2.45$\times 10^{-1}$&  4.38$\times 10^{+0}$&  5.24$\times 10^{+1}$\\
\hline
\multicolumn{7}{l}{\sc sum of the above components}  \\
\hline
I123&  8.49$\times 10^{-3}$&  7.70$\times 10^{-4}$&  2.77$\times 10^{-2}$&  8.94$\times 10^{-1}$&  9.24$\times 10^{+0}$&  1.25$\times 10^{+2}$\\
I12&  8.53$\times 10^{-3}$&  7.72$\times 10^{-4}$&  2.78$\times 10^{-2}$&  8.84$\times 10^{-1}$&  9.18$\times 10^{+0}$&  1.22$\times 10^{+2}$\\
F123&  1.62$\times 10^{-1}$&  3.94$\times 10^{-2}$&  1.99$\times 10^{-1}$&  1.92$\times 10^{+0}$&  2.83$\times 10^{+0}$&  1.14$\times 10^{+1}$\\
F12&  1.51$\times 10^{-1}$&  3.61$\times 10^{-2}$&  1.90$\times 10^{-1}$&  1.87$\times 10^{+0}$&  2.90$\times 10^{+0}$&  1.21$\times 10^{+1}$\\
I + F& 1.71$\times 10^{-1}$&  3.99$\times 10^{-2}$&  2.00$\times 10^{-1}$&  1.92$\times 10^{+0}$&  2.72$\times 10^{+0}$&  1.08$\times 10^{+1}$\\
\hline
\multicolumn{7}{l}{\sc WMAP at 94GHz}  \\
\hline
I123&  1.92$\times 10^{-2}$&  1.97$\times 10^{-3}$&  4.44$\times 10^{-2}$&  8.65$\times 10^{-1}$&  5.31$\times 10^{+0}$&  3.97$\times 10^{+1}$\\
\hline
\hline
\end{tabular}
\end{table}

% //////////// LFI9 PG31=9C ////////////
\begin{table}
\centering
\caption{The same as in Table~1, but for the indicated simulated beam pattern.}
\begin{tabular}{c c c c c c c}
\hline
\hline
\multicolumn{7}{c}{{\sc LFI9 9C}}  \\
{\sc beam}	& &	&	&	&	{\sc skewness}	&	{\sc kurtosis} \\
{\sc region}	&	{\sc average} &	{\sc variance} & {\sc rms} &	{\sc peak-to-peak} & {\sc index}	&	{\sc index} \\
\hline
\hline
\multicolumn{7}{l} {\sc dust + diffuse free--free emission}  \\
\hline
I12&  2.20$\times 10^{-2}$&  5.42$\times 10^{-3}$&  7.36$\times 10^{-2}$&  2.02$\times 10^{+0}$&  8.87$\times 10^{+0}$&  1.06$\times 10^{+2}$\\
F123&  4.71$\times 10^{-1}$&  4.17$\times 10^{-1}$&  6.46$\times 10^{-1}$&  5.84$\times 10^{+0}$&  3.00$\times 10^{+0}$&  1.24$\times 10^{+1}$\\
F12&  4.34$\times 10^{-1}$&  3.80$\times 10^{-1}$&  6.16$\times 10^{-1}$&  5.70$\times 10^{+0}$&  3.10$\times 10^{+0}$&  1.32$\times 10^{+1}$\\
I + F& 4.93$\times 10^{-1}$&  4.20$\times 10^{-1}$&  6.48$\times 10^{-1}$&  5.84$\times 10^{+0}$&  2.92$\times 10^{+0}$&  1.19$\times 10^{+1}$\\
\hline
\multicolumn{7}{l}{\sc diffuse synchrotron emission }  \\
\hline
I12&  2.04$\times 10^{-3}$&  9.95$\times 10^{-6}$&  3.15$\times 10^{-3}$&  4.59$\times 10^{-2}$&  6.65$\times 10^{+0}$&  5.76$\times 10^{+1}$\\
F123&  4.12$\times 10^{-2}$&  1.33$\times 10^{-3}$&  3.65$\times 10^{-2}$&  2.45$\times 10^{-1}$&  2.04$\times 10^{+0}$&  4.61$\times 10^{+0}$\\
F12&  3.79$\times 10^{-2}$&  1.19$\times 10^{-3}$&  3.45$\times 10^{-2}$&  2.38$\times 10^{-1}$&  2.12$\times 10^{+0}$&  5.03$\times 10^{+0}$\\
I + F& 4.32$\times 10^{-2}$&  1.33$\times 10^{-3}$&  3.65$\times 10^{-2}$&  2.44$\times 10^{-1}$&  1.98$\times 10^{+0}$&  4.39$\times 10^{+0}$\\
\hline
\multicolumn{7}{l}{\sc HII regions }  \\
\hline
I12&  7.54$\times 10^{-4}$&  5.36$\times 10^{-5}$&  7.32$\times 10^{-3}$&  5.46$\times 10^{-1}$&  2.98$\times 10^{+1}$&  1.41$\times 10^{+3}$\\
F123&  1.43$\times 10^{-2}$&  5.92$\times 10^{-4}$&  2.43$\times 10^{-2}$&  2.75$\times 10^{-1}$&  3.29$\times 10^{+0}$&  1.58$\times 10^{+1}$\\
F12&  1.31$\times 10^{-2}$&  5.37$\times 10^{-4}$&  2.32$\times 10^{-2}$&  2.59$\times 10^{-1}$&  3.39$\times 10^{+0}$&  1.68$\times 10^{+1}$\\
I + F& 1.50$\times 10^{-2}$&  6.47$\times 10^{-4}$&  2.54$\times 10^{-2}$&  5.91$\times 10^{-1}$&  3.62$\times 10^{+0}$&  2.42$\times 10^{+1}$\\
\hline
\multicolumn{7}{l}{\sc sum of the above components}  \\
\hline
I12&  2.48$\times 10^{-2}$&  6.52$\times 10^{-3}$&  8.08$\times 10^{-2}$&  2.26$\times 10^{+0}$&  8.85$\times 10^{+0}$&  1.07$\times 10^{+2}$\\
F123&  5.26$\times 10^{-1}$&  4.97$\times 10^{-1}$&  7.05$\times 10^{-1}$&  6.35$\times 10^{+0}$&  2.96$\times 10^{+0}$&  1.20$\times 10^{+1}$\\
F12&  4.85$\times 10^{-1}$&  4.52$\times 10^{-1}$&  6.72$\times 10^{-1}$&  6.18$\times 10^{+0}$&  3.05$\times 10^{+0}$&  1.27$\times 10^{+1}$\\
I + F& 5.51$\times 10^{-1}$&  5.00$\times 10^{-1}$&  7.07$\times 10^{-1}$&  6.35$\times 10^{+0}$&  2.88$\times 10^{+0}$&  1.15$\times 10^{+1}$\\
\hline
\hline
\end{tabular}
\end{table}

% //////////// LFI4 PG21=4A ////////////
\begin{table}
\centering
\caption{The same as in Table~1, but for the indicated simulated beam pattern.
For the straylight in the far sidelobes we report also the result based on 
the WMAP map at 94~GHz, including all components.}
\begin{tabular}{c c c c c c c}
\hline
\hline
\multicolumn{7}{c}{{\sc LFI4 4A}}  \\
{\sc beam}	& &	&	&	&	{\sc skewness}	&	{\sc kurtosis} \\
{\sc region}	&	{\sc average} &	{\sc variance} & {\sc rms} &	{\sc peak-to-peak} & {\sc index}	&	{\sc index} \\
\hline
\hline
\multicolumn{7}{l} {\sc dust + diffuse free--free emission}  \\
\hline
I12&  5.03$\times 10^{-4}$&  2.88$\times 10^{-6}$&  1.70$\times 10^{-3}$&  5.71$\times 10^{-2}$&  9.19$\times 10^{+0}$&  1.22$\times 10^{+2}$\\
F123&  1.42$\times 10^{-2}$&  1.83$\times 10^{-4}$&  1.35$\times 10^{-2}$&  1.02$\times 10^{-1}$&  1.73$\times 10^{+0}$&  4.00$\times 10^{+0}$\\
F12&  1.24$\times 10^{-2}$&  1.54$\times 10^{-4}$&  1.24$\times 10^{-2}$&  1.02$\times 10^{-1}$&  2.02$\times 10^{+0}$&  5.84$\times 10^{+0}$\\
I + F& 1.47$\times 10^{-2}$&  1.87$\times 10^{-4}$&  1.37$\times 10^{-2}$&  1.02$\times 10^{-1}$&  1.65$\times 10^{+0}$&  3.58$\times 10^{+0}$\\
\hline
\multicolumn{7}{l}{\sc diffuse synchrotron emission}  \\
\hline
I12&  4.62$\times 10^{-5}$&  5.13$\times 10^{-9}$&  7.16$\times 10^{-5}$&  1.07$\times 10^{-3}$&  6.72$\times 10^{+0}$&  5.96$\times 10^{+1}$\\
F123&  1.24$\times 10^{-3}$&  5.89$\times 10^{-7}$&  7.76$\times 10^{-4}$&  4.57$\times 10^{-3}$&  1.08$\times 10^{+0}$&  7.86$\times 10^{-1}$\\
F12&  1.09$\times 10^{-3}$&  4.86$\times 10^{-7}$&  6.97$\times 10^{-4}$&  4.38$\times 10^{-3}$&  1.23$\times 10^{+0}$&  1.44$\times 10^{+0}$\\
I + F& 1.29$\times 10^{-3}$&  6.02$\times 10^{-7}$&  7.76$\times 10^{-4}$&  4.56$\times 10^{-3}$&  1.02$\times 10^{+0}$&  5.69$\times 10^{-1}$\\
\hline
\multicolumn{7}{l}{\sc HII regions }  \\
\hline
I12&  1.77$\times 10^{-5}$&  2.99$\times 10^{-8}$&  1.73$\times 10^{-4}$&  1.24$\times 10^{-2}$&  3.11$\times 10^{+1}$&  1.51$\times 10^{+3}$\\
F123&  4.85$\times 10^{-4}$&  3.69$\times 10^{-7}$&  6.07$\times 10^{-4}$&  8.04$\times 10^{-3}$&  1.83$\times 10^{+0}$&  4.34$\times 10^{+0}$\\
F12&  4.19$\times 10^{-4}$&  2.97$\times 10^{-7}$&  5.45$\times 10^{-4}$&  7.74$\times 10^{-3}$&  2.09$\times 10^{+0}$&  6.13$\times 10^{+0}$\\
I + F& 5.03$\times 10^{-4}$&  4.04$\times 10^{-7}$&  6.36$\times 10^{-4}$&  1.41$\times 10^{-2}$&  2.32$\times 10^{+0}$&  1.42$\times 10^{+1}$\\
\hline
\multicolumn{7}{l}{\sc sum of the above components}  \\
\hline
I12&  5.67$\times 10^{-4}$&  3.47$\times 10^{-6}$&  1.86$\times 10^{-3}$&  6.51$\times 10^{-2}$&  9.18$\times 10^{+0}$&  1.24$\times 10^{+2}$\\
F123&  1.59$\times 10^{-2}$&  2.20$\times 10^{-4}$&  1.48$\times 10^{-2}$&  1.10$\times 10^{-1}$&  1.70$\times 10^{+0}$&  3.77$\times 10^{+0}$\\
F12&  1.39$\times 10^{-2}$&  1.85$\times 10^{-4}$&  1.36$\times 10^{-2}$&  1.11$\times 10^{-1}$&  1.97$\times 10^{+0}$&  5.52$\times 10^{+0}$\\
I + F& 1.65$\times 10^{-2}$&  2.25$\times 10^{-4}$&  1.50$\times 10^{-2}$&  1.10$\times 10^{-1}$&  1.61$\times 10^{+0}$&  3.36$\times 10^{+0}$\\
\hline
\multicolumn{7}{l}{\sc WMAP at 94~GHz}  \\
\hline
F123&  2.75$\times 10^{-2}$&  3.37$\times 10^{-4}$&  1.83$\times 10^{-2}$&  1.51$\times 10^{-1}$&  
1.33$\times 10^{+0}$&  3.00$\times 10^{+0}$\\
\hline
\hline
\end{tabular}
\end{table}

% //////////// LFI4 PG28=4B ////////////
\begin{table}
\centering
\caption{The same as in Table~1, but for the indicated simulated beam pattern.}
\begin{tabular}{c c c c c c c}
\hline
\hline
\multicolumn{7}{c}{{\sc LFI4 4B}}  \\
{\sc beam}	& &	&	&	&	{\sc skewness}	&	{\sc kurtosis} \\
{\sc region}	&	{\sc average} &	{\sc variance} & {\sc rms} &	{\sc peak-to-peak} & {\sc index}	&	{\sc index} \\
\hline
\hline
\multicolumn{7}{l} {\sc dust + diffuse free--free emission}  \\
\hline
I12&  1.28$\times 10^{-2}$&  1.97$\times 10^{-3}$&  4.44$\times 10^{-2}$&  1.46$\times 10^{+0}$&  9.80$\times 10^{+0}$&  1.42$\times 10^{+2}$\\
F12&  1.68$\times 10^{-1}$&  3.64$\times 10^{-2}$&  1.91$\times 10^{-1}$&  1.51$\times 10^{+0}$&  2.39$\times 10^{+0}$&  7.90$\times 10^{+0}$\\
I + F& 1.81$\times 10^{-1}$&  3.86$\times 10^{-2}$&  1.96$\times 10^{-1}$&  1.58$\times 10^{+0}$&  2.21$\times 10^{+0}$&  6.71$\times 10^{+0}$\\
\hline
\multicolumn{7}{l}{\sc diffuse synchrotron emission }  \\
\hline
I12&  1.17$\times 10^{-3}$&  3.33$\times 10^{-6}$&  1.82$\times 10^{-3}$&  2.80$\times 10^{-2}$&  6.80$\times 10^{+0}$&  6.11$\times 10^{+1}$\\
F12&  1.47$\times 10^{-2}$&  1.14$\times 10^{-4}$&  1.07$\times 10^{-2}$&  6.64$\times 10^{-2}$&  1.50$\times 10^{+0}$&  2.39$\times 10^{+0}$\\
I + F& 1.59$\times 10^{-2}$&  1.18$\times 10^{-4}$&  1.09$\times 10^{-2}$&  6.60$\times 10^{-2}$&  1.38$\times 10^{+0}$&  1.86$\times 10^{+0}$\\
\hline
\multicolumn{7}{l}{\sc HII regions }  \\
\hline
I12&  4.53$\times 10^{-4}$&  2.17$\times 10^{-5}$&  4.65$\times 10^{-3}$&  3.89$\times 10^{-1}$&  3.24$\times 10^{+1}$&  1.73$\times 10^{+3}$\\
F12&  5.48$\times 10^{-3}$&  6.43$\times 10^{-5}$&  8.02$\times 10^{-3}$&  6.53$\times 10^{-2}$&  2.50$\times 10^{+0}$&  8.11$\times 10^{+0}$\\
I + F& 5.93$\times 10^{-3}$&  8.78$\times 10^{-5}$&  9.37$\times 10^{-3}$&  4.11$\times 10^{-1}$&  5.89$\times 10^{+0}$&  1.25$\times 10^{+2}$\\
\hline
\multicolumn{7}{l}{\sc sum of the above components}  \\
\hline
I12&  1.44$\times 10^{-2}$&  2.38$\times 10^{-3}$&  4.88$\times 10^{-2}$&  1.65$\times 10^{+0}$&  9.81$\times 10^{+0}$&  1.45$\times 10^{+2}$\\
F12&  1.89$\times 10^{-1}$&  4.35$\times 10^{-2}$&  2.09$\times 10^{-1}$&  1.63$\times 10^{+0}$&  2.34$\times 10^{+0}$&  7.53$\times 10^{+0}$\\
I + F& 2.03$\times 10^{-1}$&  4.63$\times 10^{-2}$&  2.15$\times 10^{-1}$&  1.79$\times 10^{+0}$&  2.17$\times 10^{+0}$&  6.41$\times 10^{+0}$\\
\hline
\hline
\end{tabular}
\end{table}

% //////////// LFI4 PG29=4C ////////////
\begin{table}
\centering
\caption{The same as in Table~1, but for the indicated simulated beam pattern.}
\begin{tabular}{c c c c c c c}
\hline
\hline
\multicolumn{7}{c}{{\sc LFI4 4C}}  \\
{\sc beam}	& &	&	&	&	{\sc skewness}	&	{\sc kurtosis} \\
{\sc region}	&	{\sc average} &	{\sc variance} & {\sc rms} &	{\sc peak-to-peak} & {\sc index}	&	{\sc index} \\
\hline
\hline
\multicolumn{7}{l} {\sc dust + diffuse free--free emission}  \\
\hline
I12&  9.56$\times 10^{-3}$&  1.12$\times 10^{-3}$&  3.34$\times 10^{-2}$&  1.13$\times 10^{+0}$&  9.96$\times 10^{+0}$&  1.49$\times 10^{+2}$\\
F12&  1.08$\times 10^{-1}$&  1.64$\times 10^{-2}$&  1.28$\times 10^{-1}$&  1.03$\times 10^{+0}$&  2.55$\times 10^{+0}$&  8.77$\times 10^{+0}$\\
I + F& 1.17$\times 10^{-1}$&  1.76$\times 10^{-2}$&  1.33$\times 10^{-1}$&  1.20$\times 10^{+0}$&  2.36$\times 10^{+0}$&  7.43$\times 10^{+0}$\\
\hline
\multicolumn{7}{l}{\sc diffuse synchrotron emission }  \\
\hline
I12&  8.75$\times 10^{-4}$&  1.87$\times 10^{-6}$&  1.37$\times 10^{-3}$&  2.11$\times 10^{-2}$&  6.81$\times 10^{+0}$&  6.14$\times 10^{+1}$\\
F12&  9.43$\times 10^{-3}$&  5.01$\times 10^{-5}$&  7.08$\times 10^{-3}$&  4.49$\times 10^{-2}$&  1.64$\times 10^{+0}$&  2.90$\times 10^{+0}$\\
I + F& 1.03$\times 10^{-2}$&  5.21$\times 10^{-5}$&  7.22$\times 10^{-3}$&  4.46$\times 10^{-2}$&  1.49$\times 10^{+0}$&  2.27$\times 10^{+0}$\\
\hline
\multicolumn{7}{l}{\sc HII regions }  \\
\hline
I12&  3.40$\times 10^{-4}$&  1.24$\times 10^{-5}$&  3.53$\times 10^{-3}$&  2.93$\times 10^{-1}$&  3.24$\times 10^{+1}$&  1.71$\times 10^{+3}$\\
F12&  3.51$\times 10^{-3}$&  2.93$\times 10^{-5}$&  5.41$\times 10^{-3}$&  4.55$\times 10^{-2}$&  2.69$\times 10^{+0}$&  9.26$\times 10^{+0}$\\
I + F& 3.85$\times 10^{-3}$&  4.25$\times 10^{-5}$&  6.52$\times 10^{-3}$&  3.08$\times 10^{-1}$&  7.07$\times 10^{+0}$&  1.71$\times 10^{+2}$\\
\hline
\multicolumn{7}{l}{\sc sum of the above components}  \\
\hline
I12&  1.08$\times 10^{-2}$&  1.35$\times 10^{-3}$&  3.67$\times 10^{-2}$&  1.27$\times 10^{+0}$&  9.98$\times 10^{+0}$&  1.52$\times 10^{+2}$\\
F12&  1.21$\times 10^{-1}$&  1.96$\times 10^{-2}$&  1.40$\times 10^{-1}$&  1.12$\times 10^{+0}$&  2.50$\times 10^{+0}$&  8.38$\times 10^{+0}$\\
I + F& 1.32$\times 10^{-1}$&  2.11$\times 10^{-2}$&  1.45$\times 10^{-1}$&  1.35$\times 10^{+0}$&  2.32$\times 10^{+0}$&  7.13$\times 10^{+0}$\\
\hline
\hline
\end{tabular}
\end{table}

% //////////// LFI4 PG30=4D ////////////
\begin{table}
\centering
\caption{The same as in Table~1, but for the indicated simulated beam pattern.}
\begin{tabular}{c c c c c c c}
\hline
\hline
\multicolumn{7}{c}{{\sc LFI4 4D}}  \\
{\sc beam}	& &	&	&	&	{\sc skewness}	&	{\sc kurtosis} \\
{\sc region}	&	{\sc average} &	{\sc variance} & {\sc rms} &	{\sc peak-to-peak} & {\sc index}	&	{\sc index} \\
\hline
\hline
\multicolumn{7}{l} {\sc dust + diffuse free--free emission}  \\
\hline
I12&  1.19$\times 10^{-2}$&  1.73$\times 10^{-3}$&  4.16$\times 10^{-2}$&  1.41$\times 10^{+0}$&  9.93$\times 10^{+0}$&  1.48$\times 10^{+2}$\\
F12&  1.47$\times 10^{-1}$&  3.10$\times 10^{-2}$&  1.76$\times 10^{-1}$&  1.43$\times 10^{+0}$&  2.60$\times 10^{+0}$&  9.10$\times 10^{+0}$\\
I + F& 1.59$\times 10^{-1}$&  3.28$\times 10^{-2}$&  1.81$\times 10^{-1}$&  1.50$\times 10^{+0}$&  2.41$\times 10^{+0}$&  7.82$\times 10^{+0}$\\
\hline
\multicolumn{7}{l}{\sc diffuse synchrotron emission }  \\
\hline
I12&  1.09$\times 10^{-3}$&  2.91$\times 10^{-6}$&  1.71$\times 10^{-3}$&  2.62$\times 10^{-2}$&  6.81$\times 10^{+0}$&  6.13$\times 10^{+1}$\\
F12&  1.28$\times 10^{-2}$&  9.39$\times 10^{-5}$&  9.69$\times 10^{-3}$&  6.22$\times 10^{-2}$&  1.66$\times 10^{+0}$&  3.04$\times 10^{+0}$\\
I + F& 1.39$\times 10^{-2}$&  9.70$\times 10^{-5}$&  9.85$\times 10^{-3}$&  6.17$\times 10^{-2}$&  1.53$\times 10^{+0}$&  2.46$\times 10^{+0}$\\
\hline
\multicolumn{7}{l}{\sc HII regions }  \\
\hline
I12&  4.24$\times 10^{-4}$&  1.94$\times 10^{-5}$&  4.40$\times 10^{-3}$&  3.67$\times 10^{-1}$&  3.27$\times 10^{+1}$&  1.75$\times 10^{+3}$\\
F12&  4.80$\times 10^{-3}$&  5.55$\times 10^{-5}$&  7.45$\times 10^{-3}$&  6.36$\times 10^{-2}$&  2.73$\times 10^{+0}$&  9.59$\times 10^{+0}$\\
I + F& 5.22$\times 10^{-3}$&  7.61$\times 10^{-5}$&  8.73$\times 10^{-3}$&  3.88$\times 10^{-1}$&  6.26$\times 10^{+0}$&  1.35$\times 10^{+2}$\\
\hline
\multicolumn{7}{l}{\sc sum of the above components}  \\
\hline
I12&  1.34$\times 10^{-2}$&  2.09$\times 10^{-3}$&  4.57$\times 10^{-2}$&  1.58$\times 10^{+0}$&  9.96$\times 10^{+0}$&  1.52$\times 10^{+2}$\\
F12&  1.65$\times 10^{-1}$&  3.70$\times 10^{-2}$&  1.92$\times 10^{-1}$&  1.56$\times 10^{+0}$&  2.55$\times 10^{+0}$&  8.70$\times 10^{+0}$\\
I + F& 1.78$\times 10^{-1}$&  3.92$\times 10^{-2}$&  1.98$\times 10^{-1}$&  1.69$\times 10^{+0}$&  2.37$\times 10^{+0}$&  7.49$\times 10^{+0}$\\
\hline
\hline
\end{tabular}
\end{table}   

% TABLE SUMMARY
\begin{table}[!h]
\centering
\caption{{\sc rms} and peak-to-peak of the simulated TOD 
(antenna temperature expressed in $\mu$K) from intermediate,
far sidelobes and both of them of the all simulated beam patterns at 100~GHz 
for sum of the three considered templates of Galactic components
as functions of the edge taper. 
We neglect here the contribution from 
the third order optical interactions since 
only first and second order optical interactions have been 
evaluated for the all optical designs.}
\begin{tabular}{c c c c c c c c}
\hline
\hline
\multicolumn{8}{c}{{\sc Summary Table}}  \\
{\sc beam}      & ET &   I                &   I               &  F            &  
F               &  I + F            &  I + F \\
                & dB @ 24$^\circ$   & {\sc rms}    &  {\sc peak-to-peak}     &  {\sc rms}    &  
{\sc peak-to-peak}     &  {\sc rms}    &  {\sc peak-to-peak} \\
\hline
\hline
LFI9 9A & 25.5 &  8.38$\times 10^{-3}$&  2.94$\times 10^{-1}$ & 7.33$\times 10^{-2}$&  6.16$\times 10^{-1}$ & 7.61$\times 10^{-2}$&  6.38$\times 10^{-1}$ \\
LFI9 9B & 19.0 &  2.78$\times 10^{-2}$&  8.84$\times 10^{-1}$ & 1.90$\times 10^{-1}$&  1.87$\times 10^{+0}$ & 2.00$\times 10^{-1}$&  1.92$\times 10^{+0}$ \\
LFI9 9C & 15.0 &  8.08$\times 10^{-2}$&  2.26$\times 10^{+0}$ & 6.72$\times 10^{-1}$&  6.18$\times 10^{+0}$ & 7.07$\times 10^{-1}$&  6.35$\times 10^{+0}$ \\
LFI4 4A & 28.3 &  1.86$\times 10^{-3}$&  6.51$\times 10^{-2}$ & 1.36$\times 10^{-2}$&  1.11$\times 10^{-1}$ & 1.50$\times 10^{-2}$&  1.10$\times 10^{-1}$ \\
LFI4 4B & 19.0 &  4.88$\times 10^{-2}$&  1.65$\times 10^{+0}$ & 2.09$\times 10^{-1}$&  1.63$\times 10^{+0}$ & 2.15$\times 10^{-1}$&  1.79$\times 10^{+0}$ \\
LFI4 4C & 19.0 &  3.67$\times 10^{-2}$&  1.27$\times 10^{+0}$ & 1.40$\times 10^{-1}$&  1.12$\times 10^{+0}$ & 1.45$\times 10^{-1}$&  1.35$\times 10^{+0}$ \\
LFI4 4D & 19.0 &  4.57$\times 10^{-2}$&  1.58$\times 10^{+0}$ & 1.92$\times 10^{-1}$&  1.56$\times 10^{+0}$ & 1.98$\times 10^{-1}$&  1.69$\times 10^{+0}$ \\
\hline
\hline
\end{tabular}
\label{gsc}
\end{table}

For a couple of representative cases, LFI9 9B and LFI4 4A, we 
evaluated the straylight signal on the basis of the WMAP map
at 94~GHz for the intermediate pattern and far sidelobes, respectively
(see Tables~2 and 4).
For the signal in the intermediate pattern,
we find a straylight contamination larger by a factor $\simeq 1.6$ in terms 
of {\sc rms} (but essentially unchanged in terms of peak-to-peak).
%Previous analyses at 30~GHz (Burigana et al. 2000) 
%indicate that the relative contribution from bright point sources 
%to the straylight in the intermediate pattern is not negligible
%compared the contribution from the Galatic radio emission.
%On the other hand, for the considered optical designs at 100~GHz
%the overall straylight contamination is mainly due to the far sidelobes 
%and not to the intermediate pattern.
It is interesting to note that in the case of
the WMAP map, since the 94~GHz channel has a low level of 
Galactic foreground contamination, 
the contribution to straylight from CMB fluctuations clearly appears
(see Fig.~8) 
far from the Galactic plane, where the straylight signal 
is negligible in the case of the simulation carried out
by using the three Galactic templates
(on the contrary, this effect is not evident at 30~GHz, 
see Fig.~B.1 in Appendix~B).
If we set to zero the values of the TOD samples 
with signal smaller than $\simeq 0.15 \mu$K (the maximum value
of straylight far from the Galactic plane in the right panel of Fig.~8)
we obtain a {\sc rms} straylight of $\simeq 0.024 \mu$K,
quite close to the {\sc rms} straylight value of 
$\simeq 0.028 \mu$K obtained 
in the case of the sum of the three adopted Galactic templates
(see Table~2). 
This means that main contribution to the factor $\simeq 1.6$ 
of discrepancy in the {\sc rms} of the straylight signal 
found by adopting the three Galactic templates
or the WMAP map at 94~GHz is mainly produced by the CMB anisotropy.

In the case of the straylight in the far sidelobes
we find a straylight contamination 
larger only by a factor $\sim 1.3$ 
in the case of the WMAP map than in the case of 
the sum of the three adopted Galactic templates,
in terms of both {\sc rms} and peak-to-peak.
The above CMB anisotropy effect is not evident. On the contrary,
the regions at low straylight signal in the case of the 
simulation based on the WMAP map show a pattern that,
although with different signal values, is quite similar to that
found in the same regions by using the Galactic templates. 
This indicates that the main difference found
in this case is due to
the adopted model of Galactic templates, in agreement 
with the idea that 
CMB anisotropies (and also 
extragalactic bright sources and fluctuations, minimum 
at about 100--200~GHz) contained in the WMAP map, 
can produce only minor contaminations, 
compared to the Galactic signal, because of their smaller power 
at large angular 
scales which implies that positive and negative fluctuations 
partially compensate inside the relatively large antenna pattern solid angles 
relevant for this kind of straylight contamination. 

   \begin{figure*}
   \centering
   \begin{tabular}{ccc}
   \includegraphics[width=5cm]{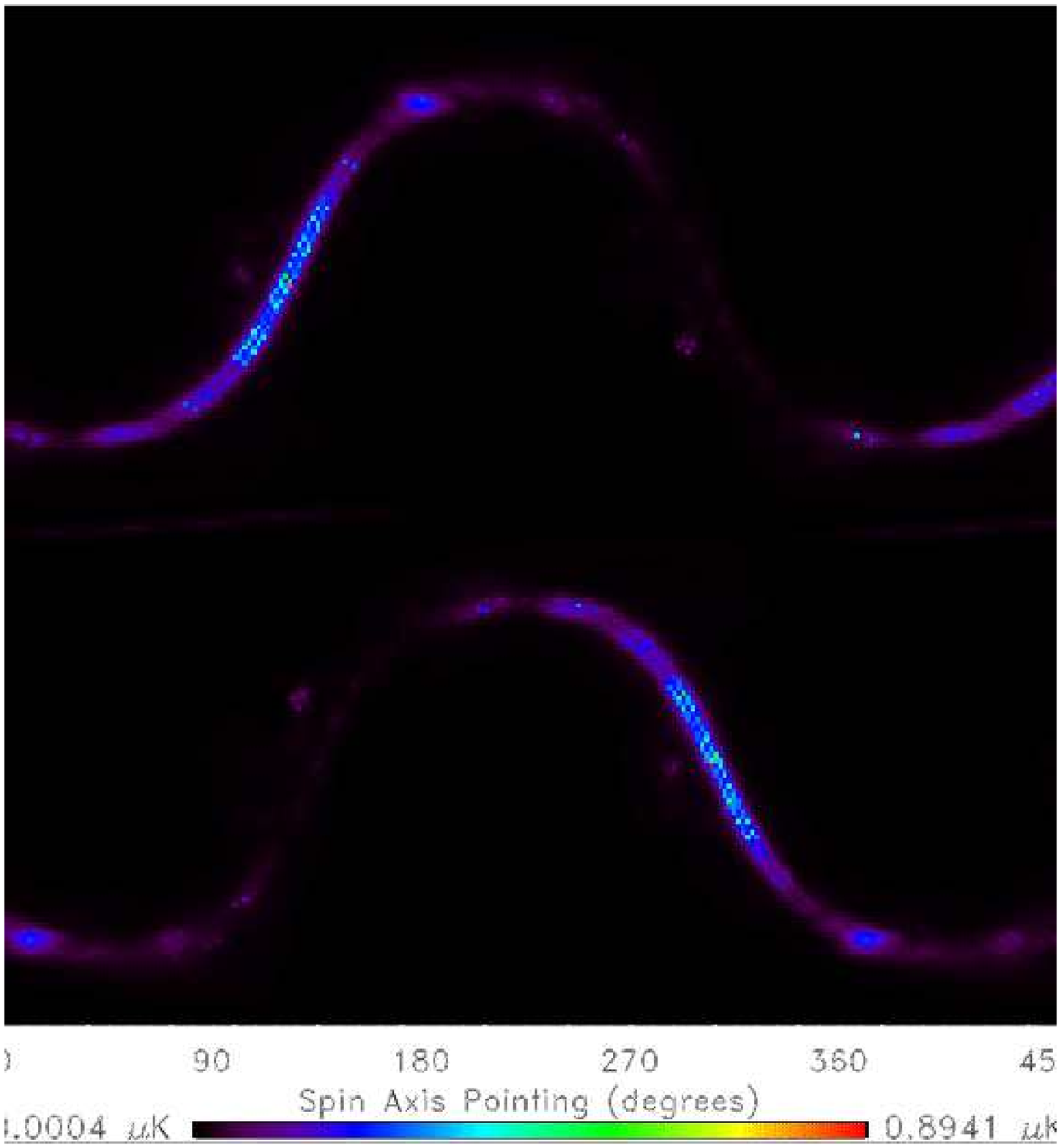}&
   \includegraphics[width=5cm]{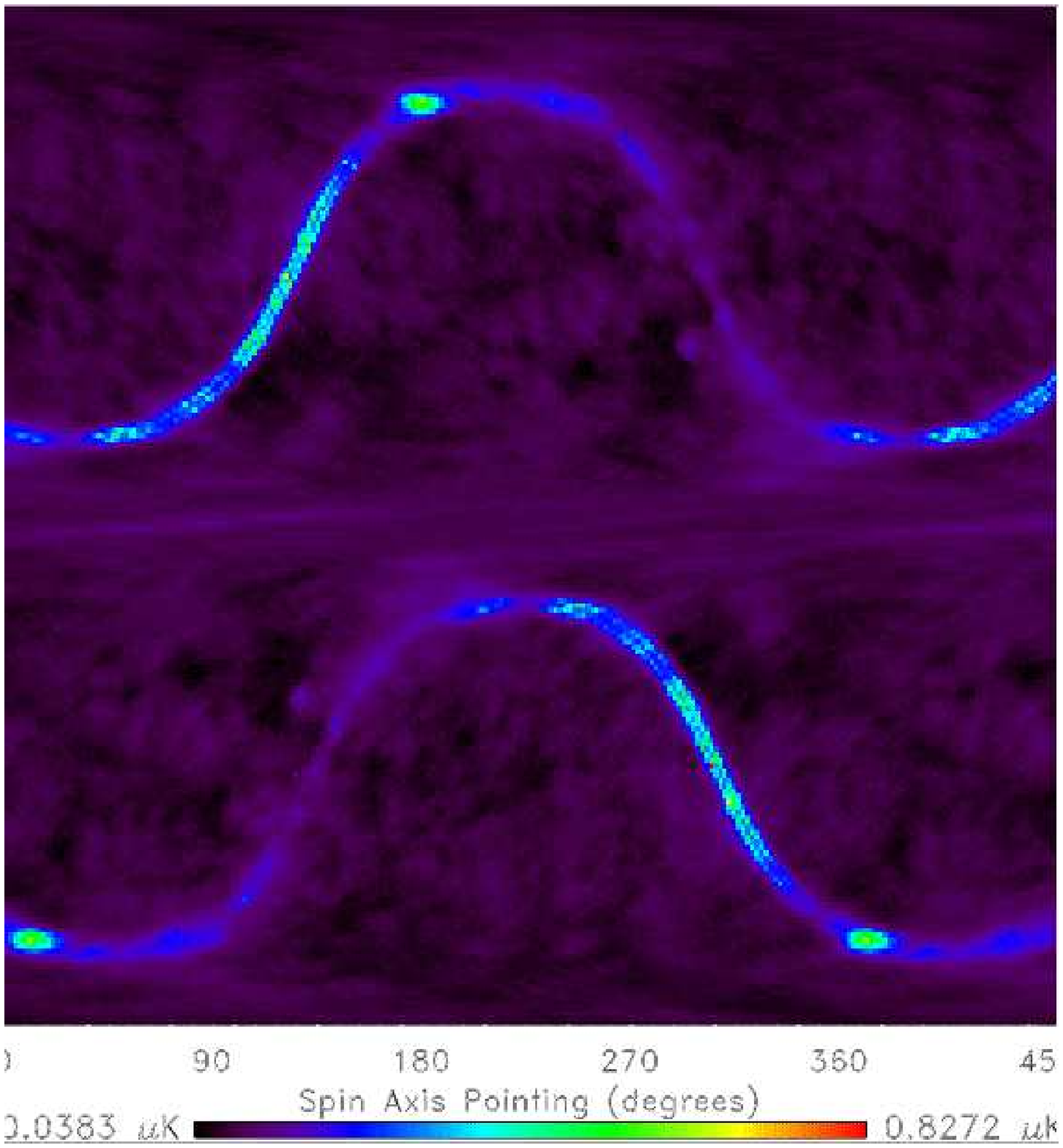}
   \end{tabular}
   \caption{The same as in Fig.~5 but for the 
sum of the three Galactic templates and for the WMAP map at
94~GHz where the contribution from CMB anisotropy 
appears (the antenna temperature in $\mu$K in linear scale
is here reported; see also the text).}
%              \label{FigGam}%
    \end{figure*}

By exploiting our set of optical configurations and taking into account
the correction factor found on the basis of the WMAP map, we
find a linear approximation describing quite well the dependence
of the {\sc rms} and the peak-to-peak values of the GSC at 100~GHz 
on the (per cent) fractional contribution, $f_\%$,  
to the integrated antenna pattern response from the considered pattern region
(see also Fig.~9):

\begin{equation}
\mbox{peak-to-peak} \simeq 5.3 \mu{\rm K} \times f_\% \simeq 175 \mu{\rm K} \times {\rm LET} 
\end{equation}

\begin{equation}
{\sc rms} \simeq 0.62 \mu{\rm K} \times f_\% \simeq 21 \mu{\rm K} \times {\rm LET} \, ,
\end{equation}

\noindent
for the far sidelobes, and

\begin{equation}
\mbox{peak-to-peak} \simeq 28 \mu{\rm K} \times f_\% \simeq 98 \mu{\rm K} \times {\rm LET} 
\end{equation}

\begin{equation}
{\sc rms} \simeq 0.86 \mu{\rm K} \times f_\% \simeq 3  \mu{\rm K} \times {\rm LET} \, ,
\end{equation}

\noindent
for the intermediate pattern~\footnote{In these fits we adopt, for uniformity, 
the numbers found by considering the first and second order optical 
interactions, available for the all cases.}, where 
eqs.~(1)--(5) of Paper~I are taken into account 
in the last equalities of these equations (the LET is 
here referred to a taper at an angle of $24^\circ$ from the feed axis).
The numerical coefficients in eq.~(6) can be multiplied by $\simeq 1.6$  
to include also the straylight contribution in the intermediate pattern 
from CMB fluctuations. 

   \begin{figure*}
   \centering
   \includegraphics[width=14cm,height=7cm]{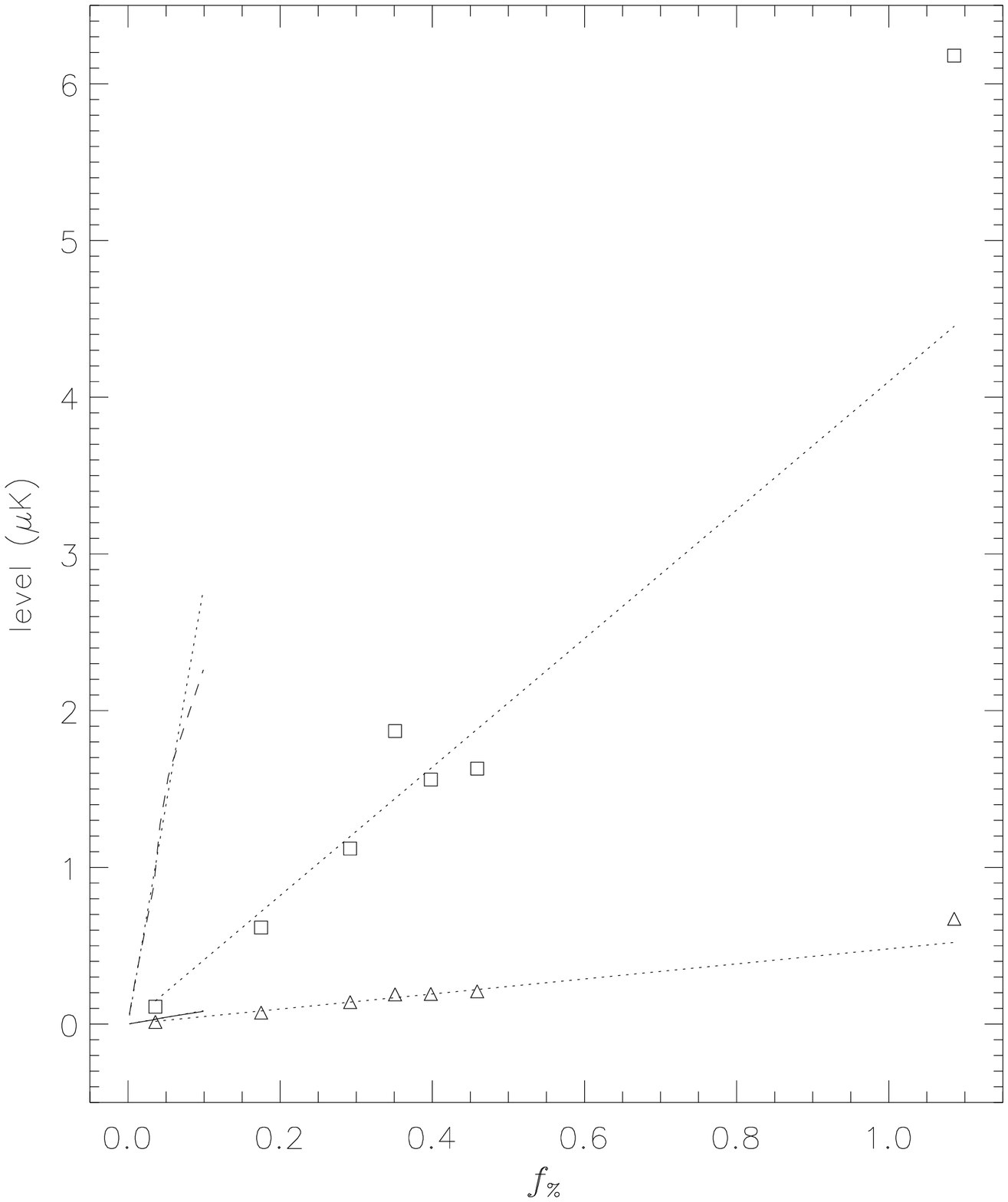}
   \caption{Dependence of the {\sc rms} (solid line for the intermediate pattern
and triangles for the far sidelobes) 
and the peak-to-peak values (dashed line for the intermediate pattern
and squares for the far sidelobes) of the GSC at 100~GHz 
on the (per cent) fractional contribution 
to the integrated antenna pattern response from the considered pattern region
and corresponding linear approximations (dotted lines).}
%              \label{FigGam}%
    \end{figure*}

Clearly, in the cosmological window the GSC contamination is less than
that in the lowest and highest {\sc Planck} frequency channels
and so its impact on CMB power spectrum recovery 
(see Burigana et al. 2001 for an analysis of GSC impact on the power spectrum
recovery and Fourier decomposition of scan circle data at the lowest 
{\sc Planck} frequency channel).
On the other hand, 
the ultimate goal of {\sc Planck}, and in general of future CMB anisotropy 
experiments after WMAP, is not only the power spectrum recovery
but also a detailed imaging of the last scattering surface and a 
detailed study of the whole statistical 
information, cosmological and astrophysical, contained in the frequency maps.
For this reason, the {\sc Planck} requirement on the maximum acceptable 
level of systematic effect contamination is of few $\mu$K,  
in order to avoid spurious signals at a level comparable with the 
{\sc Planck} sensitivity. 

Among the set of analyzed optical configurations we can identify 
a subsample which reaches a good trade-off between the angular resolution
(see Table~2 of Paper I)
and the suppression of the straylight contamination.
%We remember that the different properties of the main beam and 
%sidelobes on the antenna patterns referring to the same LFI feed position
%(e.g. LFI4 or LFI9) are related to the corresponding level of the ET
%and to the design of the feed corrugation profiles (Paper I).

In the case of LFI9 the configuration 9B shows a global 
peak-to-peak ({\sc rms}) GSC less than about $2 \mu$K ($0.2 \mu$K)
simultaneoulsy reaching an angular resolution (FWHM) of $10.02'$.
The configuration 9A shows a lower GSC, but with a worse
angular resolution ($10.56'$), while the configuration 9C
allows to reach a resolution of $9.54'$ but introducing
a relatively large GSC, being its peak-to-peak ({\sc rms}) signal of about 
$7 \mu$K ($0.7 \mu$K). 
The worse resolution for LFI4 with respect to LFI9 is due to the 
different location of the corresponding feed in the focal surface
(see Mandolesi et al. 2000b).
%Being the feed corresponding to the beam LFI4 located 
%in a worse position of the telescope focal surface for what concerns
%the main beam optical properties
%(at positive $U=x$, see Mandolesi et al. 2000b and Appendix~A),  
%its angular resolution is necessarily worse than that of LFI9
%(located at negative $U$). 
On the other hand, our analysis
identifies a well defined optimal optical configuration (4C) 
that allows to reach the best angular resolution ($12.08'$)
with the minimum GSC, i.e. a peak-to-peak ({\sc rms}) signal less than 
about $1.5 \mu$K ($0.15 \mu$K).

In Appendix B we report also the results obtained for a simulation at 30~GHz.
We find some qualitative differences between the GSC at these two 
frequencies, related to their optical behaviours and to the different role
of the Galactic foregrounds. 

An accurate computation of the GSC on {\sc Planck} polarization
data could be in principle carried out by using the formalism
described in Challinor et al. (2000).
On the other hand, since no microwave polarization surveys are currently
available, a detailed computation based on current simulated 
templates (see e.g. Giardino et al. 2002) may provide only indicative 
results. 
A first order analysis aimed to obtain a robust upper limit 
on the {\sc rms} GSC in the {\sc Planck} LFI polarization data
can be obtained with a simple argument.
The signals, $I_{ij}$ ($i=1,2$), 
in the four radiometers associated to a couple of LFI feeds
symmetrically located on the focal plane 
can be expressed as: 
$$2 I_{11}=T+\delta T^{s}_{11} 
+ (Q + \delta Q^{s}_{11}) {\rm cos}(2\phi_{11}) 
+ (U + \delta U^{s}_{11}) {\rm sin}(2\phi_{11})$$
$$2 I_{12}=T+\delta T^{s}_{12} 
- (Q + \delta Q^{s}_{12}) {\rm cos}(2\phi_{11}) 
- (U + \delta U^{s}_{12}) {\rm sin}(2\phi_{11})$$
$$2 I_{21}= T+\delta T^{s}_{21}
+ (Q + \delta Q^{s}_{21}) {\rm cos}(2\phi_{21}) 
+ (U + \delta U^{s}_{21}) {\rm sin}(2\phi_{21})$$
$$2 I_{22}= T+\delta T^{s}_{22} 
- (Q + \delta Q^{s}_{22}) {\rm cos}(2\phi_{21}) 
- (U + \delta U^{s}_{22}) {\rm sin}(2\phi_{21}) \, .$$
Here $\phi_{21}=\phi_{11}+\pi/4$, $\phi_{11}$ is the angle 
between the axis $x_{bf}$ of the {\it beam frame} corresponding
to the first feed and the 
direction of the parallel in the considered pointing direction,
$T$, $Q$ and $U$ are the unpolarized signal and 
the Stokes parameters in the main beam,
and the terms $\delta T^{s}_{ij}, \delta Q^{s}_{ij}, \delta Q^{s}_{ij}$
account for the spurious contributions due to the Galactic straylight.
For each pointing direction, they are different for the four 
radiometers because of the different level and orientation 
of the corresponding antenna patterns.
The measure of the Stokes parameters $Q$ and $U$ is obtained
by combining the signals in the four radiometers:
$Q=(I_{11}-I_{12}) {\rm cos}(2\phi_{11}) - (I_{21}-I_{22}) {\rm sin}(2\phi_{11})$, 
$U=(I_{21}-I_{22}) {\rm cos}(2\phi_{11}) + (I_{11}-I_{22}) {\rm sin}(2\phi_{11})$.
The Galactic synchrotron emission is partially
polarized ($\sim 30$~\%), while the Galactic
free-free and thermal dust emission, most relevant at 
$\nu \gsim 50$~GHz, are only weakly polarized (a few \%).
Therefore, at least in the cosmological window, 
a Galactic straylight contamination relevant for the 
polarization measure mainly derives from the differences 
between the temperature straylight signals, $\delta T^{s}_{ij}$,  
in each pairs of radiometers associated to the same feed.
Assuming a typical value $\sim 1/2$ for ${\rm cos}(2\phi_{11})$ and 
${\rm sin}(2\phi_{11})$, we find a {\sc rms} GSC on $Q$ and $U$ 
similar to that found above for $T$ [see eqs. (4) and (6)],
to be considered as a pessimistic upper limit, corresponding 
to a difference of a factor $\simeq 2$ 
in the temperature straylight signals
in each pair of radiometers.

We then conclude that, at least in terms of {\sc rms} and in the 
cosmological window, keeping at very low level the GSC 
in the temperature data directly assures an adequate suppression of 
the GSC in the polarization data.

\section{Conclusions}

Satellite CMB anisotropy missions, such as WMAP
and {\sc Planck}, and also the new generation of balloon-borne and 
ground experiments, make use of complex multi-frequency instruments
at the focal surface of a meter class telescope to allow the 
joint study of CMB and foreground anisotropies, necessary for a
high quality component separation. In the so-called ``cosmological
window'', between $\sim 70$~GHz and $\sim 300$~GHz, where 
foreground contamination is minimum, it is extremely 
important to reach the best angular resolution 
(necessary to measure the high order acoustic peaks of CMB anisotropy) 
achievable keeping at the same time the straylight contamination 
at acceptable levels (peak-to-peak of few $\mu$K).

By focusing, as a working case, on the 100~GHz channels 
of {\sc Planck} LFI,
we have presented here extensive simulations of the straylight contamination
starting from a wide set of simulated optical 
configurations, described in Paper~ I, in order to find
the best compromise between resolution and GSC. 

By adopting some templates of Galactic foreground extrapolated from 
radio and IR surveys 
we found that it is possible 
to improve the angular resolution of about $5-7$~\%
and to reach for example $10'-12'$ of FWHM at 100~GHz
by keeping 
the overall straylight contamination below the 
level of few $\mu$K in terms of peak-to-peak and about 10 times smaller in terms
of {\sc rms}, as necessary to avoid systematic errors comparable with
the {\sc Planck} sensitivity. 

We compared the level of straylight introduced by the 
different Galactic components for different beam regions 
and provided simple approximate relations giving the {\sc rms} and 
peak-to-peak levels of the GSC for the intermediate pattern 
and the far sidelobes as functions of the corresponding contributions 
to the integrated antenna pattern response, 
related to the edge taper.

For the considered 
optical designs, the most important straylight contamination 
derives from the far sidelobes, where the Galactic signal 
overwhelms the other straylight contributions
(nevertheless, in the intermediate pattern, the straylight contamination 
from CMB fluctuations is found to be not negligible compared to the GSC).

We demonstrated that including the third order optical interactions
changes only of few \% the results of 
straylight analyses. As discussed in Paper~I, this is extremely
important for optical design optimization studies, probing
that accurate enough optical simulations can be carried
out by saving about 75\% of the computational time 
without a relevant loss of accuracy.

Keeping at very low level the GSC 
in the temperature data directly assures 
a reliable suppression of 
the GSC in the polarization data,
at least in terms of {\sc rms}.
The CMB polarization angular  
power spectrum ($ET$, $E$, and $B$ modes) recovery
will be then not significantly affected by straylight.
 
The results at 100~GHz have been compared 
with those at 30~GHz, where the GSC is more critical,
showing a peak-to-peak value at a level of $\simeq 5-7\mu$K. 
In comparison with 
previous analyses (see e.g. Burigana et al. 2001), this represents 
a relevant improvement in straylight rejection related to the 
optimization of the overall {\sc Planck} optical scheme 
(Dubruel et al. 2000).
Clearly, assuming different Galactic templates implies 
differences in the computed straylight signals. 
On the other hand, even for the quite different input maps
adopted in these tests, 
differences larger than $\simeq 1-2\mu$K
are limited to quite localized regions, close to the 
Galactic plane, where the CMB anisotropy is dominated 
by the very high Galactic signal in the main beam.
This suggests that, even 
far from the ``cosmological window'',
a subtraction of the GSC well down to $\sim 1\mu$K level 
could be obtained by evaluating with few iterations the straylight 
contamination signal for the maps directly derived from
{\sc Planck} observations, provided that 
the antenna pattern response could be 
quite accurately modelled.

\begin{acknowledgements}
%We warmly thank our colleagues of the {\sc Planck} Consortia
%for numberless and helpful discussions.
Some of the results in this paper have been derived using the HEALPix
(G\`orski et al. 1999).
%We gratefully acknowledge K.M.~G\'orski and all the people
%involved in the realization of the tools of HEALPix pixelisation.
%We wish to thank the referee for constructive comments.

\end{acknowledgements}

\appendix

\section{Transformation rules between {\it telescope frame} and {\it beam frame}}

Let $\vec s$ be the unit vector, choosen outward the Sun direction,
of the spin axis direction and 
$\hat{k}$ that of the direction, $z$, of the telescope line of sight (LOS),
pointing at an angle $\alpha \sim 85^{\circ}$ from the direction of $\vec s$. 

On the plane tangent to the celestial 
sphere in the direction of the LOS
we choose two coordinates $x$ and $y$, respectively defined by the unit vector
$\hat{i}$ and $\hat{j}$
according to the convention that the unit vector
$\hat{i}$ points always toward $\vec s$
and that $x,y,z$ is a standard Cartesian frame,
referred here as {\it telescope frame}.

Let $\hat{i}_{bf},\hat{j}_{bf},\hat{k}_{bf}$ be
the unit vectors corresponding to the Cartesian axes 
$x_{bf},y_{bf},z_{bf}$ of the {\it beam frame}; $\hat{k}_{bf}$ defines the  
direction of the beam centre axis in the {\it telescope frame}.
The {\it beam frame} is defined with respect to the {\it telescope frame}
by three angles: $\theta_B$,~$\phi_B$,~$\psi_B$ ($\theta_B$ and $\phi_B$,
two standard polar coordinates defining the direction of the 
beam centre axis, range respectively from $0^\circ$, for an on-axis beam, to some
degrees, for LFI off-axis beams, and from $0^\circ$ to $360^\circ$).

Let $\hat{i'}_{bf},\hat{j'}_{bf},\hat{k'}_{bf'}$ 
($\hat{k'}_{bf}=\hat{k}_{bf}$) be the unit vectors corresponding to 
the Cartesian axes $x',y',z'$ of an {\it intermediate frame}, defined
by the two angles $\theta_B$ and $\phi_B$, obtained by
the {\it telescope frame} $x,y,z$ when the unit vector of the axis $z$ 
is rotated by an angle $\theta_B$ on the plane defined by the unit vector 
of the axis $z$ and the unit vector $\hat{k}_{bf}$ up to reach $\hat{k}_{bf}$:

%        beamvec(0)=cos(phibeamrad)*sin(thbeamrad)
%        beamvec(1)=sin(phibeamrad)*sin(thbeamrad)
%        beamvec(2)=cos(thbeamrad)
\begin{equation}
\hat{k'}_{bf} = \hat{k}_{bf} =  {\rm cos}(\phi_B) {\rm sin}(\theta_B) \hat{i}
              + {\rm sin}(\phi_B) {\rm sin}(\theta_B) \hat{j} 
              + {\rm cos}(\theta_B) \hat{k}
\end{equation}

\begin{equation}
\hat{i'}_{bf} =  [{\rm cos}(\phi_B)^2 {\rm cos}(\theta_B) + {\rm sin}(\phi_B)^2] \hat{i}
              + [{\rm sin}(\phi_B) {\rm cos}(\phi_B) ({\rm cos}(\theta_B)-1)] \hat{j}
              - {\rm sin}(\theta_B) {\rm cos}(\phi_B) \hat{k} 
\end{equation}

\begin{equation}
\hat{j'}_{bf} =  [{\rm sin}(\phi_B) {\rm cos}(\phi_B) ({\rm cos}(\theta_B)-1)] \hat{i}
              + [{\rm cos}(\theta_B) {\rm sin}(\phi_B)^2 + {\rm cos}(\phi_B)^2] \hat{j}
              - {\rm sin}(\theta_B) {\rm sin}(\phi_B) \hat{k} \, . 
\end{equation}

%        i_mb_inframe_pm(0)=cos(phibeamrad)**2*cos(thbeamrad)
%     &                     +sin(phibeamrad)**2
%        i_mb_inframe_pm(1)=sin(phibeamrad)*cos(phibeamrad)
%     &                     *(cos(thbeamrad)-1.d+0)
%        i_mb_inframe_pm(2)=-sin(thbeamrad)*cos(phibeamrad)
%
%        j_mb_inframe_pm(0)=i_mb_inframe_pm(1)
%        j_mb_inframe_pm(1)=cos(thbeamrad)*sin(phibeamrad)**2
%     &                     +cos(phibeamrad)**2
%        j_mb_inframe_pm(2)=-sin(thbeamrad)*sin(phibeamrad)

The {\it beam frame} is obtained from the {\it intermediate frame}
through a further (anti-clockwise) rotation of an angle $\psi_B$ 
(ranging from $0^\circ$ to 
$360^\circ$~\footnote{We note that, in other conventions, angles $\phi'_B$ and $\psi'_B$
ranging from $-180^\circ$ to $180^\circ$ are given, instead of $\phi_B$ and $\psi_B$. 
The angles $\phi_B$ and $\psi_B$ here defined are 
equal to $\phi'_B$ and $\psi'_B$ when they are positive 
and are given respectively by 
$360^\circ + \phi'_B$ and $360^\circ + \psi'_B$ for negative $\phi'_B$ and $\psi'_B$.})
around $\hat{k}_{bf}$ and is therefore
explicitely given by:

\begin{eqnarray}
\hat{i}_{bf} & = &  [{\rm cos}(\psi_B) \hat{i'}_{bf,x} + {\rm sin}(\psi_B) \hat{j'}_{bf,x}] \hat{i}   
              +  [{\rm cos}(\psi_B) \hat{i'}_{bf,y} + {\rm sin}(\psi_B) \hat{j'}_{bf,y}] \hat{j} \nonumber \\
             & ~ &  +  [{\rm cos}(\psi_B) \hat{i'}_{bf,z} + {\rm sin}(\psi_B) \hat{j'}_{bf,z}] \hat{z}
\end{eqnarray}

%        ii_mb_inframe_pm(0)=cos(psibeamrad)*i_mb_inframe_pm(0)
%     &                      +sin(psibeamrad)*j_mb_inframe_pm(0)
%        ii_mb_inframe_pm(1)=cos(psibeamrad)*i_mb_inframe_pm(1)
%     &                      +sin(psibeamrad)*j_mb_inframe_pm(1)
%        ii_mb_inframe_pm(2)=cos(psibeamrad)*i_mb_inframe_pm(2)
%     &                      +sin(psibeamrad)*j_mb_inframe_pm(2)

\begin{eqnarray}
\hat{j}_{bf} & = & [-{\rm sin}(\psi_B) \hat{i'}_{bf,x} + {\rm cos}(\psi_B) \hat{j'}_{bf,x}] \hat{i}
              + [-{\rm sin}(\psi_B) \hat{i'}_{bf,y} + {\rm cos}(\psi_B) \hat{j'}_{bf,y}] \hat{j} \nonumber \\
             & ~ &  + [-{\rm sin}(\psi_B) \hat{i'}_{bf,z} + {\rm cos}(\psi_B) \hat{j'}_{bf,z}] \hat{z} \, ,
\end{eqnarray}

%        jj_mb_inframe_pm(0)=-sin(psibeamrad)*i_mb_inframe_pm(0)
%     &                      +cos(psibeamrad)*j_mb_inframe_pm(0)
%        jj_mb_inframe_pm(1)=-sin(psibeamrad)*i_mb_inframe_pm(1)
%     &                      +cos(psibeamrad)*j_mb_inframe_pm(1)
%        jj_mb_inframe_pm(2)=-sin(psibeamrad)*i_mb_inframe_pm(2)
%     &                      +cos(psibeamrad)*j_mb_inframe_pm(2)

\noindent
where the bottom index $x$ $(y,z)$ indicates the component of 
{\it intermediate frame} unit vector along 
the axis $x$ $(y,z)$ of the {\it telescope frame}, as defined by 
eqs.~(A1--A3).

\section{Simulations at 30~GHz}

The feed horn at 30 GHz considered in these simulations is specified by its
Spherical Wave Expansion (SWE) provided by Alcatel Space Industries, since
the sub reflector is in the near field of the corrugated horn and near
field effects cannot be neglected. The feed horn directivity is about 22
dBi, the ET is 30 dB at $22^\circ$, and the main beam has a FWHM resolution of 
$33.73'$. The beam position and orientation is identified by
($\theta_B$,~$\phi_B$,~$\psi_B$) = ($4.3466^\circ$,~$153.6074^\circ$,~$337.5^\circ$).

We carry out the simulation of {\sc Planck} observation as described in Sect.~2
by assuming both the Galactic templates presented in Sect.~2.2 and the WMAP map
at 33~GHz. The results are summarized in Table~7 while
Fig.~B.1 reports the TOD corresponding to the overall straylight
signal (from the far sidelobes plus intermediate pattern) 
for the sum of the three Galactic components described in Sect.~2.2 
and for the WMAP map. In spite of the differences in the foreground 
templates (the dominant signal deriving from Galactic diffuse free-free 
emission by using the templates of Sect.~2.2 and, according to 
Bennett et al. 2003b,
from Galactic diffuse synchrotron emission by using the  WMAP map)
the peak-to-peak values is at a level of $\simeq 5-7\mu$K for the 
contributions from  both the far sidelobes and the intermediate pattern
({\sc rms}~$\sim 1\mu$K, mainly due to the signal in the far sidelobes).
Right panel of Fig.~B.1 reports shows the difference between
the TOD obtained by using these two different templates:
only for $\simeq 0.01$~\% (0.09, 0.3, 10~\%) of the samples
of the TOD the difference is larger than $4\mu$K (3, 2, $1\mu$K).
The figure shows also that differences larger than $\simeq 1-2\mu$K
are localized quite close to the Galactic plane, where the 
CMB anisotropy is dominated by the very high Galactic 
signal in the main beam.
This implies that a subtraction of the GSC well down to $\sim 1\mu$K level
does not require a particularly accurate 
description of the microwave sky emission
nor particularly sophisticated computations,
even at frequencies where the GSC is relevant, but it is mainly 
related to a good knowledge of the antenna pattern response.

Clearly, the unsubtracted GSC is relevant 
at frequencies far from the ``cosmological window''. On the other hand,
this results represent a significant optical improvement compared to the 
analyses of GSC at 30~GHz by Burigana et al. (2001), based on 
the optical simulations by De~Maagt et al. (1998), 
predicting a similar level of GSC from the far sidelobes but 
a contamination from the intermediate pattern significantly worst 
(peak-to-peak of about $15\mu$K). 
This reduction of GSC from the intermediate pattern
is due to the corresponding contribution to the integrated response 
from the antenna pattern at few degrees from the beam centre direction, 
significantly reduced in the actual optimized optical 
design~\footnote{
In fact, the (per cent) fractional contribution, $f_\%$, is 
now $\simeq 0.045$ (0.39) 
for the intermediate pattern (far sidelobes), while it were
about 0.6 (1) for the intermediate pattern (far sidelobes)
in the case of the previous analysis.}.

   \begin{figure*}
   \centering
   \begin{tabular}{ccc}
   \includegraphics[width=5cm]{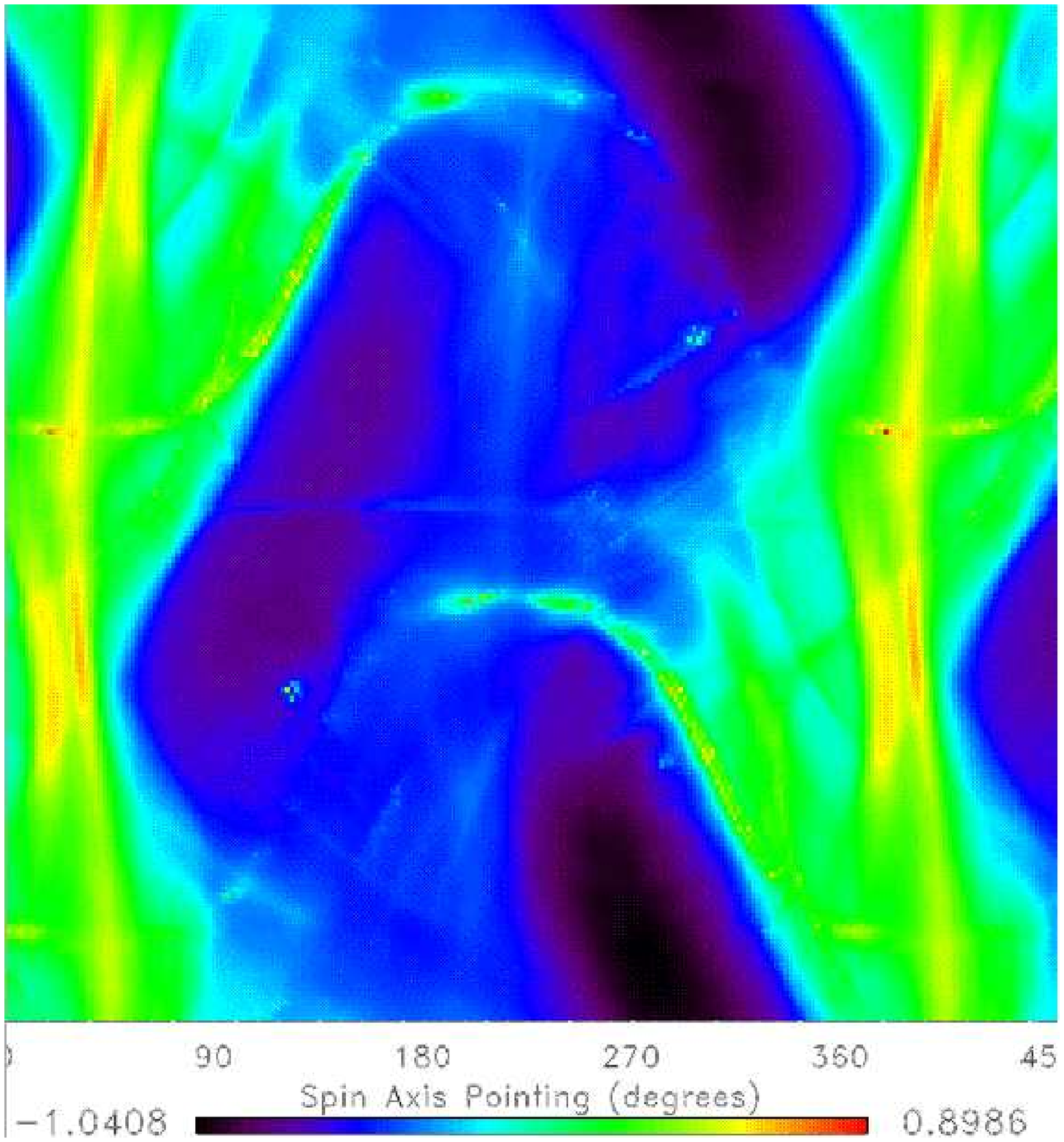}&
   \includegraphics[width=5cm]{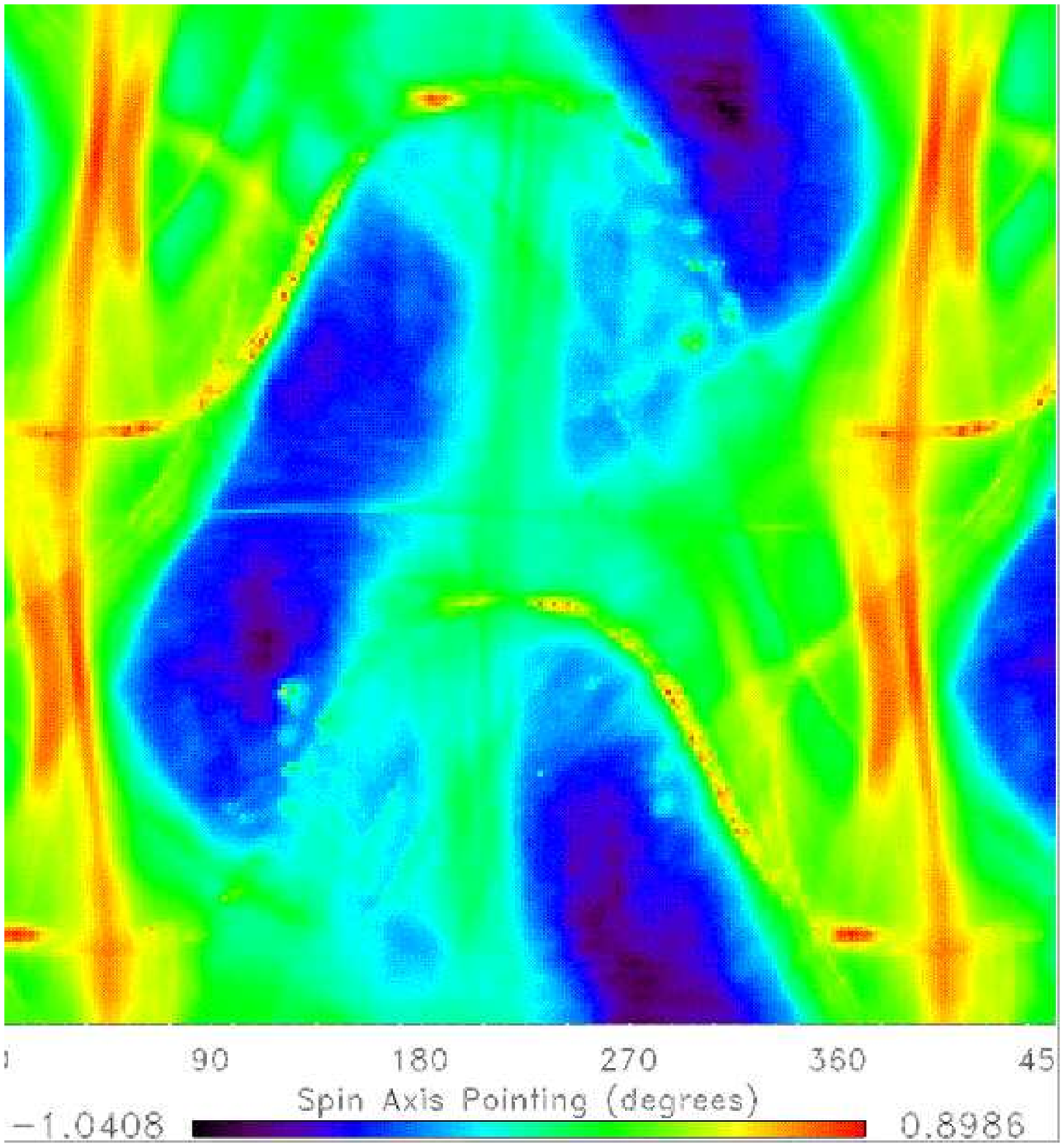}&
   \includegraphics[width=5cm]{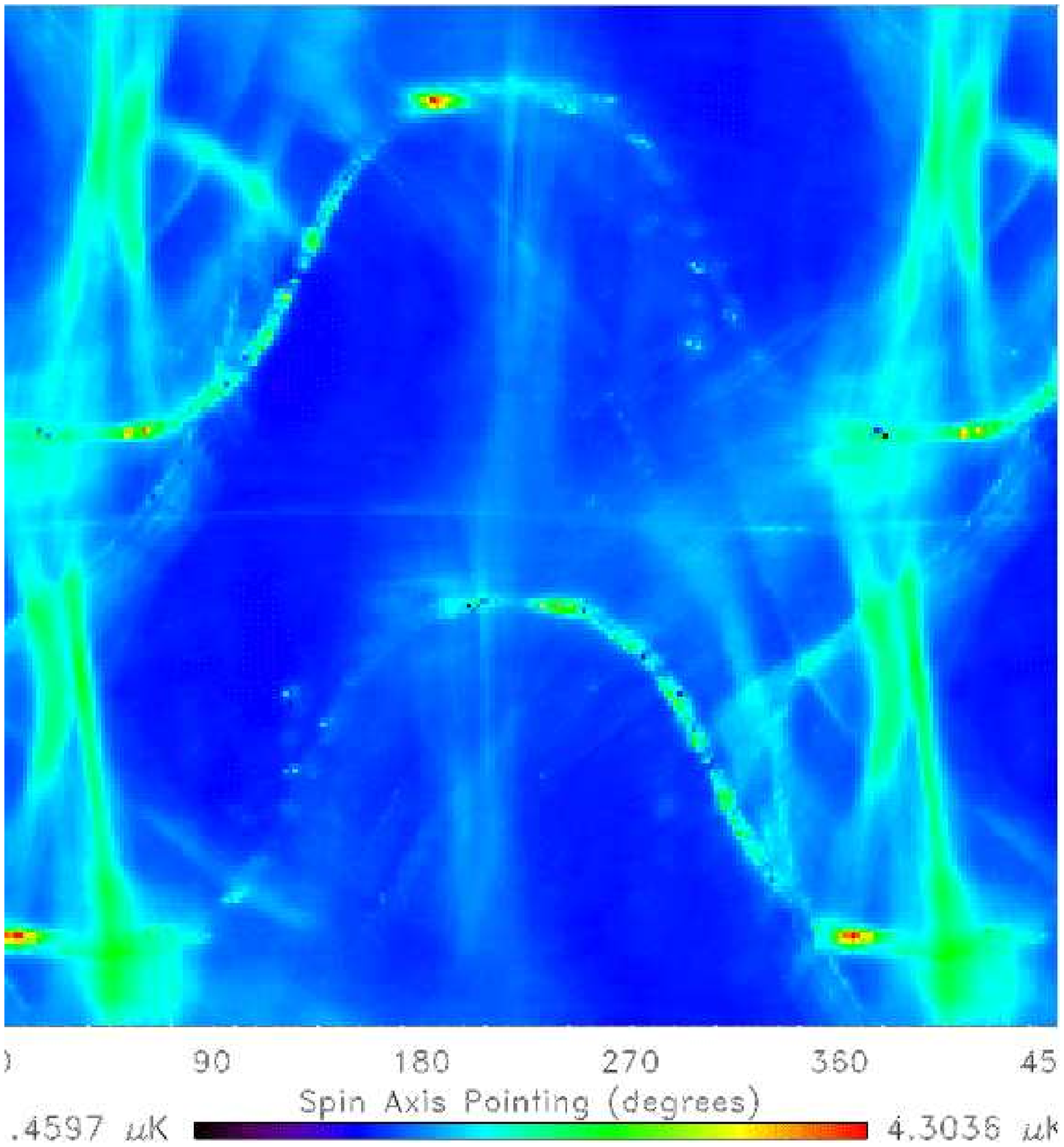}
   \end{tabular}
   \caption{The same as in Fig.~4, but for the simulation at 30~GHz
and the overall straylight signal. 
For a better comparison the adopted temperature range 
is the same in the two panels, although the mimimum and maximum values
are just different, as evident from the different peak-to-peak values
reported in Table~B.1. The right panel reports
(antenna temperature in linear, not logarithm, scale, in this case)
the difference between the middle panel and the left panel (see also the text).}
%              \label{FigGam}%
    \end{figure*}

% //////////// LFI27 SWE ////////////
\begin{table}
\centering
\caption{The same as in Table~1, but for the beam pattern simulated at 30~GHz.
We report also the result based on 
the WMAP map at 33~GHz, including all components.}
\begin{tabular}{c c c c c c c}
\hline
\hline
\multicolumn{7}{c}{{\sc LFI27 SWE}}  \\
{\sc beam}	& &	&	&	&	{\sc skewness}	&	{\sc kurtosis} \\
{\sc region}	&	{\sc average} &	{\sc variance} & {\sc rms} &	{\sc peak-to-peak} & {\sc index}	&	{\sc index} \\
\hline
\hline
\multicolumn{7}{l} {\sc dust + diffuse free--free emission}  \\
\hline
I12&  3.13$\times 10^{-2}$&  1.15$\times 10^{-2}$&  1.07$\times 10^{-1}$&  2.72$\times 10^{+0}$&  9.45$\times 10^{+0}$&  1.21$\times 10^{+2}$\\
F12&  4.75$\times 10^{-1}$&  2.62$\times 10^{-1}$&  5.12$\times 10^{-1}$&  3.09$\times 10^{+0}$&  1.64$\times 10^{+0}$&  2.32$\times 10^{+0}$\\
I + F& 5.07$\times 10^{-1}$&  2.71$\times 10^{-1}$&  5.20$\times 10^{-1}$&  3.09$\times 10^{+0}$&  1.53$\times 10^{+0}$&  1.90$\times 10^{+0}$\\
\hline
\multicolumn{7}{l}{\sc diffuse synchrotron emission }  \\
\hline
I12&  2.97$\times 10^{-2}$&  2.16$\times 10^{-3}$&  4.64$\times 10^{-2}$&  7.22$\times 10^{-1}$&  6.72$\times 10^{+0}$&  5.89$\times 10^{+1}$\\
F12&  4.26$\times 10^{-1}$&  9.94$\times 10^{-2}$&  3.15$\times 10^{-1}$&  1.47$\times 10^{+0}$&  1.27$\times 10^{+0}$&  7.35$\times 10^{-1}$\\
I + F& 4.56$\times 10^{-1}$&  1.01$\times 10^{-1}$&  3.17$\times 10^{-1}$&  1.46$\times 10^{+0}$&  1.18$\times 10^{+0}$&  4.90$\times 10^{-1}$\\
\hline
\multicolumn{7}{l}{\sc HII regions }  \\
\hline
I12&  4.33$\times 10^{-3}$&  2.54$\times 10^{-3}$&  5.04$\times 10^{-2}$&  4.65$\times 10^{+0}$&  4.27$\times 10^{+1}$&  2.91$\times 10^{+3}$\\
F12&  5.79$\times 10^{-2}$&  5.78$\times 10^{-3}$&  7.60$\times 10^{-2}$&  7.61$\times 10^{-1}$&  1.80$\times 10^{+0}$&  3.73$\times 10^{+0}$\\
I + F& 6.23$\times 10^{-2}$&  8.40$\times 10^{-3}$&  9.17$\times 10^{-2}$&  4.84$\times 10^{+0}$&  8.41$\times 10^{+0}$&  2.84$\times 10^{+2}$\\
\hline
\multicolumn{7}{l}{\sc sum of the above components}  \\
\hline
I12&  6.54$\times 10^{-2}$&  3.18$\times 10^{-2}$&  1.78$\times 10^{-1}$&  5.34$\times 10^{+0}$&  9.35$\times 10^{+0}$&  1.25$\times 10^{+2}$\\
F12&  9.60$\times 10^{-1}$&  8.02$\times 10^{-1}$&  8.96$\times 10^{-1}$&  5.05$\times 10^{+0}$&  1.51$\times 10^{+0}$&  1.74$\times 10^{+0}$\\
I + F& 1.03$\times 10^{+0}$&  8.28$\times 10^{-1}$&  9.10$\times 10^{-1}$&  7.77$\times 10^{+0}$&  1.42$\times 10^{+0}$&  1.46$\times 10^{+0}$\\
\hline
\multicolumn{7}{l}{\sc WMAP at 33~GHz}  \\
\hline
I12&  1.10$\times 10^{-1}$&  1.05$\times 10^{-1}$&  3.24$\times 10^{-1}$&  6.17$\times 10^{+0}$&  7.75$\times 10^{+0}$&  7.72$\times 10^{+1}$\\
F12&  1.37$\times 10^{+0}$&  1.51$\times 10^{+0}$&  1.23$\times 10^{+0}$&  6.76$\times 10^{+0}$&  1.40$\times 10^{+0}$&  1.45$\times 10^{+0}$\\
I + F& 1.48$\times 10^{+0}$&  1.61$\times 10^{+0}$&  1.27$\times 10^{+0}$&  7.18$\times 10^{+0}$&  1.29$\times 10^{+0}$&  1.07$\times 10^{+0}$\\
\hline
\hline
\end{tabular}
\end{table}

\end{document}